\documentclass[sigconf, nonacm]{acmart}
\AtBeginDocument{%
  }

\usepackage{booktabs}
\usepackage{multirow}
\usepackage{enumitem}
\usepackage{makecell}
\usepackage{xspace}
\usepackage{adjustbox}
\usepackage{wrapfig}

\usepackage{tikz}
\usetikzlibrary{arrows.meta,positioning,calc,fit}
\usepackage{fontawesome5}
\usepackage{subcaption}

\usepackage{graphicx}
\usepackage{hyperref}

\usepackage{listings}
\usepackage{xcolor}
\usepackage{caption}    %
\usepackage{fvextra}
\RecustomVerbatimEnvironment{verbatim}{Verbatim}{
  breaklines,
  breakanywhere,
  frame=leftline,
  framerule=1.2pt,
  rulecolor=\color{black!25},
  framesep=0.8em,
  xleftmargin=0.25em
}
\usepackage{algorithm}
\usepackage{algpseudocode}

\newif\ifdraft
\draftfalse %

\ifdraft
  \usepackage{xcolor}
  \newcommand{\notes}[2]{{\bf\textsf{\textcolor{#1}{#2}}}}
\else
  \newcommand{\notes}[2]{}
\fi

\newcommand{\eat}[1]{}
\usepackage{letltxmacro}
\usepackage{fvextra}
\fvset{breaklines=true,breakanywhere=true}
\LetLtxMacro{\originalverb}{\verb}
\LetLtxMacro{\verb}{\Verb}

\newcommand{\toolname}{\mbox{\textsc{LLMVD.js}}\xspace}
\newcommand{\nodejs}{\mbox{Node.js}\xspace}
\newcommand{\npm}{npm\xspace}
\newcommand{\nodemedic}{\textsc{NodeMedic-FINE}\xspace}
\newcommand{\fasttool}{\textsc{FAST}\xspace}
\newcommand{\explodejs}{\mbox{\textsc{Explode.js}}\xspace}

\usepackage{listings}
\lstdefinelanguage{JavaScript}{
  keywords={typeof, new, true, false, catch, function, return, null, catch, switch, var, if, in, while, do, else, case, break, const, await, async, yield, eval, exec, spawn},
  keywordstyle=\color{blue},
  ndkeywords={class, export, boolean, throw, implements, import, this},
  ndkeywordstyle=\color{darkgray},
  identifierstyle=\color{black},
  sensitive=false,
  comment=[l]{//},
  morecomment=[s]{/*}{*/},
  commentstyle=\color{olive},
  stringstyle=\color{purple},
  morestring=[b]',
  morestring=[b]"
}
\newcommand{\annotationcolor}{\color{green}}
\lstdefinestyle{customjs}{
  language=JavaScript,
  extendedchars=true,
  basicstyle=\scriptsize\ttfamily,
  showstringspaces=false,
  showspaces=false,
  numbers=left,
  numberstyle=\footnotesize,
  tabsize=2,
  breaklines=false,
  showtabs=false,
  captionpos=b,
  frame=single,
  framerule=0.2pt,
  morecomment=[l][\annotationcolor\bfseries]{\#},
  framexleftmargin=2em,
  xleftmargin=2.5em,
  xrightmargin=0.5em,
}
\lstdefinelanguage{Python}{
  keywords={true, false, except, def, return, if, elif, in, while, else, yield, for},
  keywordstyle=\color{blue},
  ndkeywords={class, self},
  ndkeywordstyle=\color{darkgray},
  identifierstyle=\color{black},
  sensitive=false,
  comment=[l]{\#},
  morecomment=[s]{"""}{"""},
  commentstyle=\color{olive},
  stringstyle=\color{purple},
  morestring=[b]',
  morestring=[b]"
}
\lstdefinestyle{custompy}{
  language=Python,
  extendedchars=true,
  basicstyle=\scriptsize\ttfamily,
  showstringspaces=false,
  showspaces=false,
  numbers=left,
  numberstyle=\footnotesize,
  tabsize=2,
  breaklines=false,
  showtabs=false,
  captionpos=b,
  frame=single,
  framerule=0.2pt,
  framexleftmargin=2.5em,
  xleftmargin=2.5em,
  xrightmargin=0.5em,
}

\newcommand{\js}[1]{\lstinline[style=customjs]{#1}\xspace}

\newcommand{\RNparagraph}[1]{\vspace{5pt}\noindent{\bf #1.}}

\acmConference[]{}{}{}

\renewcommand\footnotetextcopyrightpermission[1]{}

\begin{document}

\title{Taint-Style Vulnerability Detection and Confirmation for \nodejs Packages Using LLM Agent Reasoning}

\author{Ronghao Ni}
\email{ronghaon@andrew.cmu.edu}
\affiliation{%
  \institution{Carnegie Mellon University}
  \city{}
  \state{}
  \country{}
}

\author{Mihai Christodorescu}
\email{christodorescu@google.com}
\affiliation{%
  \institution{Google}
  \city{}
  \state{}
  \country{}
}

\author{Limin Jia}
\email{liminjia@andrew.cmu.edu}
\affiliation{%
  \institution{Carnegie Mellon University}
  \city{}
  \state{}
  \country{}
}

\begin{abstract}
The rapidly evolving \nodejs ecosystem 
currently includes millions of 
packages and is a critical part of modern software supply chains, making
vulnerability detection of \nodejs packages increasingly important. However, traditional program analysis struggles in this setting because of dynamic JavaScript features 
and the large number of package dependencies. Recent advances in large language models (LLMs) and the emerging paradigm of LLM-based agents offer an alternative to handcrafted program models. This raises the question of whether an LLM-centric, tool-augmented approach can effectively detect and confirm taint-style vulnerabilities (e.g., arbitrary command injection) in \nodejs packages. We implement \toolname, a multi-stage agent pipeline to scan code, propose vulnerabilities, generate proof-of-concept exploits, and validate them through lightweight execution oracles; and systematically evaluate 
its effectiveness in taint-style vulnerability detection and confirmation in \nodejs packages without dedicated static/dynamic analysis engines for path derivation.  
For packages from public benchmarks, 
\toolname confirms 84\% of the vulnerabilities, compared to less than 22\% for prior program analysis tools. It also outperforms a prior LLM--program-analysis hybrid approach while requiring neither vulnerability annotations nor prior vulnerability reports. When evaluated on a set of 260 recently released packages (without vulnerability groundtruth information),
traditional tools produce validated exploits for few ($\leq 2$) packages, while \toolname generates validated exploits for 36 packages.

\end{abstract}

\keywords{Automatic Vulnerability Detection, \nodejs, Large Language Models, ReAct Agents, LLM Agents, Exploit Generation, Vulnerability Confirmation}

\maketitle
\begingroup
\renewcommand{\thefootnote}{}
\footnote{Preprint.}
\addtocounter{footnote}{-1}
\endgroup

\section{Introduction}

The \nodejs ecosystem comprises millions of JavaScript packages and is among the most widely used software platforms today. However, numerous studies have shown that a substantial fraction of these packages contain security vulnerabilities~\cite{zimmermann2019small,decan2018impact} and have been exploited in software supply-chain attacks~\cite{kula2018developers,ohm2020backstabber}. %
To ensure application security, it is critical to be able to identify vulnerabilities within this ecosystem
and 
in recent years researchers have proposed %
tools to automatically detect and confirm vulnerabilities in \nodejs packages~\cite{fast,explode,nodemedic-fine,nodemedic}. These tools primarily target taint-style vulnerabilities, including OS command injection, code injection, prototype pollution, and path traversal. %
These tools use a wide range of analysis techniques, such as dynamic taint analysis, code-property-graph-based static analysis, symbolic execution, and constraint-based synthesis. 

While successful in uncovering many real-world vulnerabilities, these tools share several fundamental challenges stemming from inherent limitations of the underlying analysis techniques.
First, JavaScript's highly dynamic nature makes accurate modeling of language semantics difficult. The lack of precise type information further complicates analysis. Second, \nodejs native (built-in) functions are implemented in C++, requiring either instrumentation of the V8 engine or manually constructed %
abstractions. Third, these tools %
 rely on external JavaScript analysis infrastructure, such as parsers~\cite{esprima}, transpilers~\cite{babel}, and instrumentation tools~\cite{sen2013jalangi}. These dependencies are inherently brittle due to JavaScript's complex language features and evolving standards. 
Finally, many approaches depend on satisfiability modulo theories (SMT) solvers, which often fail to handle constraints involving string operations and regular-expression matching.

Large Language Models (LLMs) have shown strong performance on coding-related tasks, including code generation and code comprehension~\cite{chen2021evaluating}.
Rather than relying on purpose-fit abstractions, LLMs leverage extensive pre-trained knowledge and reasoning abilities to comprehend complex code patterns and dynamic program behaviors.
Furthermore, LLM agents can iteratively refine their outputs by incorporating feedback from previous unsuccessful attempts.
These capabilities suggest LLMs may be able to overcome the limitations of traditional program analysis techniques.

This paper aims to answer the following question: \textit{Can an LLM-centric, tool-augmented workflow effectively detect and confirm vulnerabilities in \npm packages without dedicated static/dynamic analysis engines?} To answer this question, we design and implement \toolname, a ReAct-based~\cite{yao2022react} agent that leverages large language models to perform taint-style vulnerability detection and confirmation for \nodejs packages.
We evaluate \toolname on three datasets (existing public benchmarks, one private benchmark, and a set of recently released \npm packages) and show that it confirms 84\% of public benchmark vulnerabilities with valid exploits, substantially outperforming prior program-analysis tools and also outperforming an LLM+program-analysis hybrid system~\cite{pocgen} 
 while requiring significantly less prior information, and further discovers 36 vulnerabilities in recently released packages.

For the rest of this paper, we use {\em rule-based program analysis} to refer to analysis techniques such as symbolic execution and taint analysis that do not rely on machine learning components. We use {\em LLM-centric reasoning} to refer to LLM-agent reasoning over raw source code with lightweight tooling (e.g., search, execution, and oracles), but without dedicated static/dynamic analysis engines for taint/path derivation.
 
Our contributions are as follows:
\begin{itemize}
\item A systematic evaluation of a multi-stage ReAct-style LLM agent framework for taint-style vulnerability detection and confirmation in \nodejs packages, with direct comparison against state-of-the-art program-analysis tools and a program-analysis--aided LLM approach.
\item A multi-dataset evaluation setup for LLM-agent vulnerability research: combining standard public benchmarks, transformed benchmark variants for memorization robustness checks, a private real-world dataset without CVEs/public exploits, and recently released \npm packages to assess generalizability under realistic settings.
\item \toolname identified 36 previously undocumented vulnerabilities in recently released \nodejs packages. 
\end{itemize}

\noindent{\bf Ethical Considerations}
Our work raises inherent dual-use concerns due to \toolname's ability to automatically detect %
vulnerabilities and 
generate exploits; however, we believe that the defensive benefits outweigh the associated risks. %
All experiments were conducted in sandboxed environments; no production systems or external servers were targeted. We analyze only %
open-source packages. We reported all 36 previously unreported validated vulnerabilities identified in newly released packages to maintainers and have received acknowledgments from 3 maintainers.

\section{Background and Related Work}
Vulnerability detection is the task of identifying security-relevant flaws in software that may be exploited by adversaries~\cite{chess2004static,landwehr1994taxonomy,pistoia2007survey, qian2025software}.
Vulnerability detection tools typically only report potential vulnerabilities and a separate confirmation step is needed to remove false positives. To reduce the costly manual confirmation effort, researchers have developed automated vulnerability confirmation methods that synthesize proof-of-concept (PoC) exploits~\cite{nodemedic,nodemedic-fine,explode,avgerinos2014automatic}. 
We review most recent work on vulnerability detection and confirmation of \nodejs packages and applying LLMs to vulnerability detection.

\subsection{\nodejs Vulnerability Detection and Confirmation}
\label{subsec:background-nodejs}

\paragraph{\nodejs Taint-style Vulnerability Detection.}
Recent vulnerability detection tools for \nodejs packages~\cite{fast,explode,nodemedic,nodemedic-fine,ntantogian2021nodexp,kim2022dapp} focus on taint-style vulnerabilities, partly because they have easy to detect code patterns and partly because they can lead to serious consequences such as allowing attackers to inject arbitrary code or execute arbitrary commands. 
For detection, these tools need to identify tainted paths from attacker controlled inputs to arguments of a sink. In the case of command injection, code injection, and path traversal, the sinks are known APIs such as \js{exec}, \js{Function}, \js{File.Write}. For prototype pollution, the vulnerability pattern involves two tainted paths and specific object field accesses. 

To compare against LLM-centric agent reasoning without dedicated static/dynamic analysis engines, we use \fasttool~\cite{fast}, \nodemedic~\cite{nodemedic-fine}, and \explodejs~\cite{explode} as representative examples of rule-based program analysis tools.
 \fasttool and \explodejs use code-property-graph (CPG) based methods for detection. They generate (their own custom) graphs representing information such as dependency, object relationship, and key operations; then vulnerability detection is reduced to a graph query. \nodemedic on the other hand, instruments JavaScript at the source level to implement dynamic taint tracking for vulnerability detection. The current implementation of \nodemedic only detects command injection and code injection vulnerabilities.  

\begin{figure}
  \begin{minipage}[t]{0.49\textwidth}
    \captionsetup{type=lstlisting}
    \captionof{lstlisting}{Code snippet of a vulnerable API}
    \label{lst:arpping_api}
    \begin{lstlisting}[style=customjs,basicstyle=\small\ttfamily,breaklines=true,breakatwhitespace=false,columns=fullflexible]
const { exec } = require('child_process');
Arpping.prototype.ping = function(range) {
    ...
    return new Promise((resolve, reject) => {
        range.forEach(ip => {
            exec(`ping ${flag} ${this.timeout} ${ip}`, (err, stdout, stderr) => {
                ...
    });});});
}
    \end{lstlisting}
  \end{minipage}
  \hfill
  \begin{minipage}[t]{0.49\textwidth}
    \captionsetup{type=lstlisting}
    \captionof{lstlisting}{PoC exploit}
        \label{lst:arpping_poc}
    \begin{lstlisting}[style=customjs,basicstyle=\small\ttfamily,breaklines=true,breakatwhitespace=false,columns=fullflexible]
const Arpping = require('./index');
(async () => {
    const a = new Arpping({ timeout: 1 });
    const payload = ['127.0.0.1; touch /tmp/os_cmd_success'];
    await a.ping(payload);
})();
    \end{lstlisting}
  \end{minipage}
    \caption{The vulnerable \npm package \texttt{arpping@2.0.0} (Snyk ID: SNYK-JS-ARPPING-1060047).}
  \label{fig:arpping_example}
\end{figure}

\paragraph{Vulnerability confirmation via exploit generation}
To generate PoC exploits, the tools need to generate a driver (testing harness) that can trigger a call to the vulnerable API (sink) and find inputs that can not only reach the sink, but also deliver the desired attack payload (e.g., a command of attacker's choice for command injection vulnerabilities). 
For example, \autoref{lst:arpping_api} shows one vulnerable API in the \texttt{arpping} package (Snyk ID: SNYK-JS-ARPPING-1060047). The \js{ping} function constructs a command string using attacker-controlled input \js{range} and passes it to the \js{exec} function, leading to an OS command injection vulnerability. To confirm this vulnerability, the tool needs to generate a driver (\autoref{lst:arpping_poc}) that calls the \js{ping} function with an input that injects an attacker-controlled command.

For inputs, \fasttool and \explodejs use symbolic execution to identify path constraints to reach the vulnerable API.
\nodemedic synthesizes input constraints from the taint provenance graph, which is an output from the dynamic taint analysis and documents all the operations that the tainted inputs underwent before reaching the sink. 
Only \explodejs is capable of generating drivers that can chain multiple API calls to reach the sink. It does so by querying its custom code property graph to identify a linear call chain. 
\nodemedic and \fasttool, instead, use a fix template to directly call the vulnerable API. 
All tools rely on SMT solvers to resolve constraints.

\subsection{LLMs in Software Vulnerability Detection}

\paragraph{LLM-assisted Security Audit.}
Recent work has explored using LLMs as assistants for security auditing, either as standalone vulnerability detectors or as components that augment traditional program analysis workflows~\cite{zhu2025software,sheng2025llms,xu2024large,jelodar2025large}. 
Several studies systematically evaluate the capability of LLMs on vulnerability detection and code analysis tasks. For example, Fang et al.\ analyze the strengths and failure modes of LLMs for code reasoning in security-relevant settings~\cite{fang2024large}, while Lin et al.\ conduct a large-scale comparative evaluation of LLM configurations across multiple datasets and programming languages~\cite{lin2025large}.
Similar evaluation efforts further characterize prompt sensitivity, model scale effects, and generalization limits in vulnerability detection~\cite{pearce2025asleep,nie2025vulnllm}. 
LLMs have been increasingly integrated into end-to-end auditing pipelines that combine detection, testing, and repair. Prior work demonstrates LLM-guided fuzzing and protocol testing, where models infer grammars and message sequences to improve coverage for stateful implementations~\cite{meng2024large, wang2024prophetfuzz}. Other systems leverage LLMs to assist patch generation and automated repair under realistic constraints~\cite{nong2025appatch, fan2023automated}.

LLMs have also been integrated into \nodejs vulnerability analysis~\cite{li2025safegenbench,hannan2025selecting,le2024study}. 
Most closely related to our work, PoCGen utilizes LLMs alongside static and dynamic analyses to interpret vulnerability reports, draft candidate exploits, and iteratively validate and refine them for \npm vulnerabilities~\cite{pocgen}. PoCGen relies on CodeQL-based static taint analysis to identify candidate vulnerable functions and input-to-sink paths, and employs dynamic analysis by executing generated exploits in a sandboxed \nodejs environment with vulnerability-specific runtime oracles to validate exploit success and guide refinement. 
PoCGen demonstrates the effectiveness of tightly integrating LLMs with program analysis and argues that plain LLM-based agents are insufficient for this task~\cite{pocgen}; in contrast, we investigate whether a carefully designed LLM reasoning–only agent can achieve competitive performance. Accordingly, we include PoCGen as a baseline in our evaluation.

\paragraph{LLM Agents in Security Auditing}
A parallel line of work studies LLMs as \emph{agents} that can plan, use tools, and iteratively refine hypotheses, moving beyond single-pass vulnerability labeling toward multi-step auditing~\cite{gasmi2025bridging,henke2025autopentest,zhu2024teams,shen2025pentestagent}. PentestGPT formalizes an LLM-driven penetration-testing workflow with modular agent roles that decompose high-level objectives into actionable testing steps and tool interactions%
~\cite{deng2024pentestgpt}. More broadly, recent studies demonstrate that agentic LLM systems can exploit real-world known vulnerabilities from vulnerability descriptions, highlighting both the potential and the risks of autonomous offensive capability when paired with tool use and structured memory~\cite{fang2024llm}. 
However, to our knowledge, no prior work has studied the performance and limitations of a carefully designed LLM agent for end-to-end taint-style vulnerability detection and confirmation in \nodejs packages, which is the focus of this paper.

\section{Motivation}

In this section, we discuss fundamental challenges that rule-based program analysis tools face when detecting and confirming taint-style vulnerabilities in \nodejs packages and outline why the capabilities of LLMs in a ReAct-based agent design framework suit this task.

\begin{table}[t!]
  \centering
  \caption{Limitations of three representative program-analysis-based tools for \nodejs taint-style vulnerability detection and confirmation:
 \fasttool~\cite{fast}, \nodemedic~\cite{nodemedic-fine},
    \explodejs~\cite{explode}} 
  \label{tab:pa_tool_limitations}
  \setlength{\tabcolsep}{6pt}
  \renewcommand{\arraystretch}{1.15}
  \begin{adjustbox}{max width=\columnwidth}
  \begin{tabular}{{p{0.02\linewidth}p{0.25\linewidth}p{0.71\linewidth}p{0.02\linewidth}}}
    \toprule
    \multicolumn{2}{l}{\textbf{Challenges}} &
    \multicolumn{2}{p{0.75\columnwidth}}{{\bf Why challenging}}
    \\
     \midrule
  \textbf{C1} &
    Generating drivers for PoC &
 \multicolumn{2}{p{0.73\columnwidth}}{Drivers may include complex interactions that make multiple API calls, set up global environment correctly, and construct and invoke call-backs.}  
\\
\cmidrule{2-4}
\multirow{3}*{\rotatebox[origin=c]{270}{Tools}} 
&  \fasttool 
& Does not generate drivers. &\multirow{3}*{\rotatebox[origin=c]{270}{Limits}}
\\ &  \nodemedic & Uses fixed driver templates and cannot handle complex interactions.  
\\ & \explodejs& Supports only linear call chains &
\\
\midrule
\textbf{C2} &
  Hard to analyze code units &
  \multicolumn{2}{p{0.73\columnwidth}}{Imported dependencies significantly increase the complexity of the analysis. Native operations are managed internally by the JavaScript engine and therefore need delicate custom handling.}
\\ \cmidrule{2-4}
\multirow{3}*{\rotatebox[origin=c]{270}{Tools}}  
 & \fasttool &  Needed manual modeling of native functions. Current implementation is missing significant ($> 90\%$) support. &\multirow{3}*{\rotatebox[origin=c]{270}{Limits}}
\\ & \nodemedic  & Applies over-approximated tainting policy.
\\ & \explodejs & %
Needs manually crafted symbolic summaries for imported APIs, resulting in low detection rate in \nodejs packages in the wild.
\\
\midrule
\textbf{C3} &
  Lacking type info. &
  \multicolumn{2}{p{0.73\columnwidth}}{Analysis needs arguments of the correct type and object structure.} 
\\ \cmidrule{2-4}
\multirow{3}*{\rotatebox[origin=c]{270}{Tools}} 
  & \fasttool &  Ignore types &\multirow{3}*{\rotatebox[origin=c]{270}{Limits}}
\\ & \nodemedic   & Algorithms for reconstructing types, neither sound nor complete
\\ & \explodejs & Querying the CPG to reconstruct types, neither sound nor complete
\\
\midrule
\textbf{C4} &
 Reliant on other JavaScript analysis infrastructure
   & 
  \multicolumn{2}{p{0.73\columnwidth}}{Analysis tools need another set of complex tools such as parsers, transpilers, and instrumentation tools, to implement their custom analysis. These tools are brittle due to JavaScript's standard evolutions and complex features.}  
\\ \cmidrule{2-4}
\multirow{3}*{\rotatebox[origin=c]{270}{Tools}} 
  & \fasttool & Esprima~\cite{esprima} parsing errors and call-edge issues. 
  &\multirow{3}*{\rotatebox[origin=c]{270}{Limits}}
  \\ & \nodemedic   & Jalangi2's~\cite{sen2013jalangi} lack of support for ES6+ 
  \\ & \explodejs & Graph.js~\cite{graphjs} exits with errors for many packages. %
 \\
\midrule
\textbf{C5} & Reliant on SMT & 
 \multicolumn{2}{p{0.73\columnwidth}}{The analysis generates constraints on string operations and regular expression matching, which are difficult to handle for SMT solvers and is an active area of research. } 
\\ \cmidrule{2-4}
\multirow{3}*{\rotatebox[origin=c]{270}{Tools}} 
  & \fasttool 
   & Z3~\cite{de2008z3} timeout when solving path constraints. 
   &\multirow{3}*{\rotatebox[origin=c]{270}{Limits}}
  \\ & \nodemedic & SMT solver timeout when solving constraints on input. 
  \\  & \explodejs & SMT solver timeout when solving path and input constraints.  
\\ \bottomrule
  \end{tabular}
\end{adjustbox}
\end{table}

\subsection{Challenges in Rule-based Program Analysis}

Recall from \autoref{subsec:background-nodejs} state-of-the-art rule-based tools~\cite{fast,explode,nodemedic-fine} implement dynamic taint tracking and code-property-graph-based static analysis for detection; and leverages symbolic execution and constraint-based synthesis for generating inputs for PoCs. Some use~\cite{explode} CPG-based static analysis for generating drivers that include complex code patterns (i.e., not directly call the vulnerable API). 
Despite substantial improvements, these tools continue to face challenges arising from the highly dynamic nature of JavaScript, the ongoing evolution of the JavaScript language, large and complex dependencies, and their reliance on other sophisticated analysis infrastructures.
These challenges are common across existing tools and stem from the fundamental limitations of the underlying techniques, rather than from limitations of individual implementation choices.
We summarize these challenges in \autoref{tab:pa_tool_limitations} and explain how each tool partially addresses them.
In practice, these limitations often lead to missed vulnerabilities or failures in exploit confirmation.
Although these tools will continue to improve, as long as the same core techniques are employed, progress is likely to be incremental~\cite{vulcan,brito2023study}. Alternatively, tools may be increasingly specialized, exploring different trade-off spaces to achieve high efficiency for specific vulnerability classes or to target different application domains (e.g., Mini apps~\cite{wang2023taintmini,zhang2024minicat}, React web apps~\cite{guo2024reactappscan}, and Electron apps~\cite{jin2023security,ali2024rise}).

\subsection{Advantages of LLM Agents}

\begin{figure}[t]
  \centering
  \begin{subfigure}{0.75\linewidth}
    \centering
    \includegraphics[width=\linewidth]{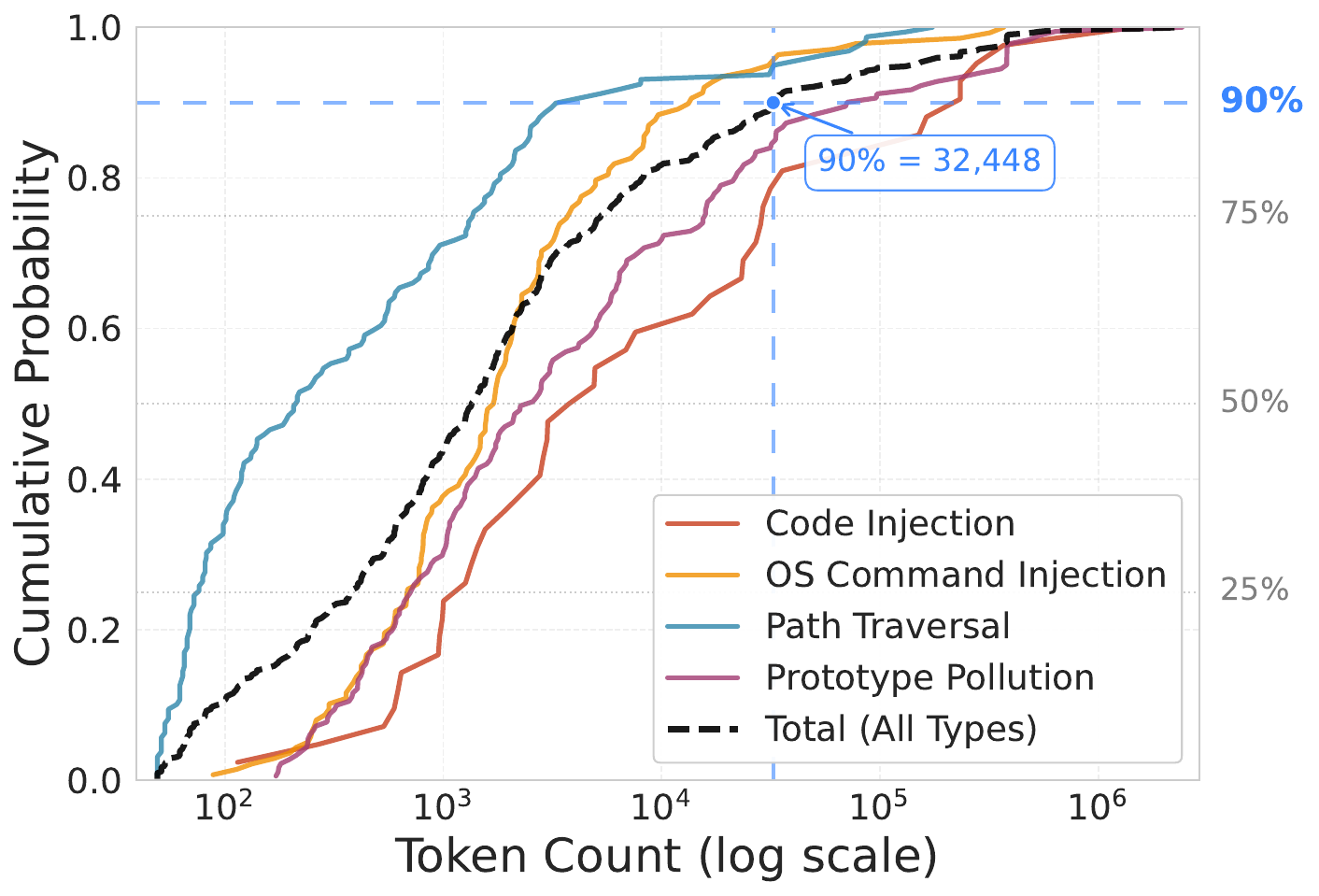}
    \caption{SecBench.js \& VulcaN datasets}
  \end{subfigure}
  \\\vspace{4pt}
  \begin{subfigure}{0.75\linewidth}
    \centering
    \includegraphics[width=\linewidth]{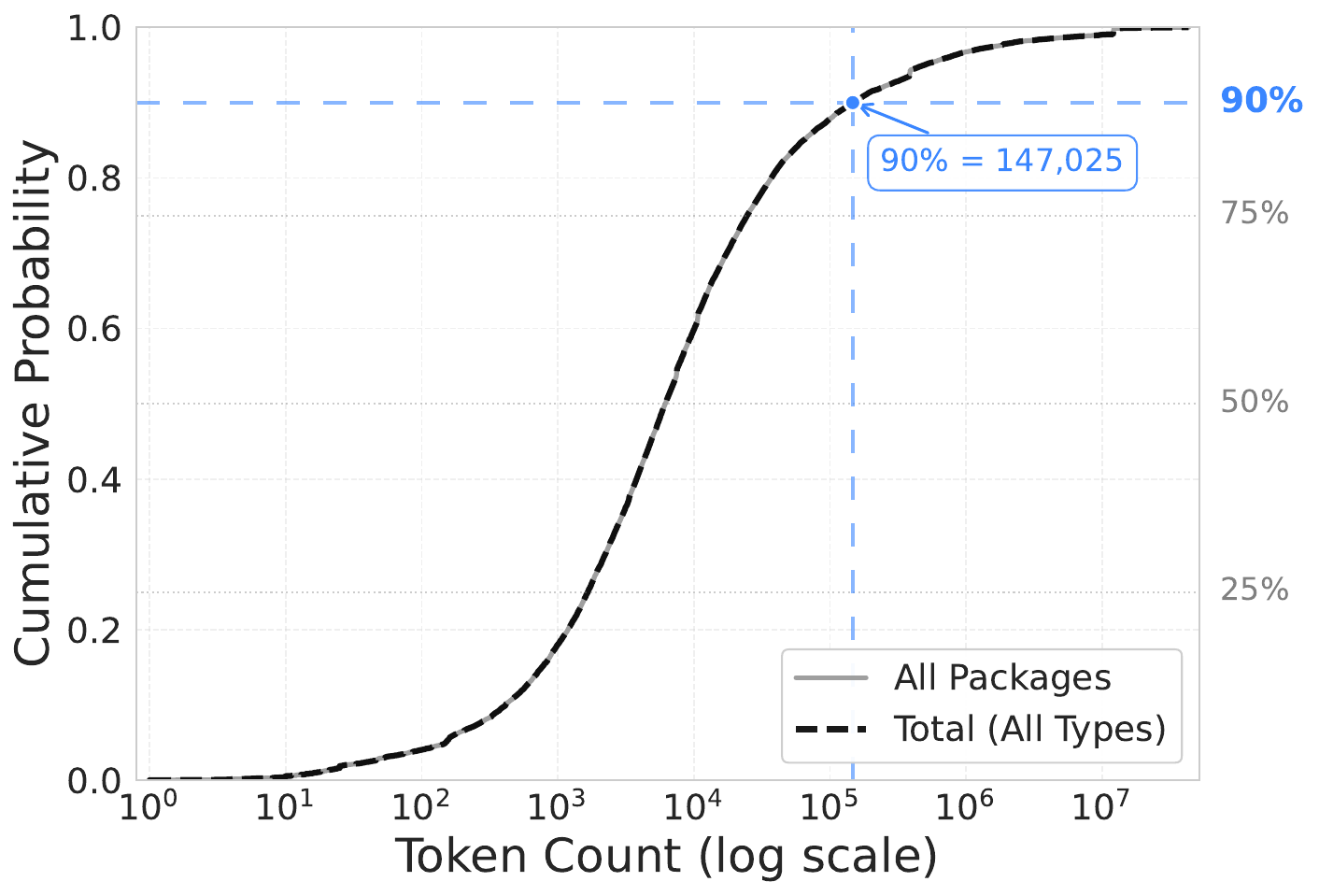}
    \caption{Recently crawled \npm packages (17,151 packages)}
  \end{subfigure}
  \caption{Cumulative distribution function (CDF) of token counts using the gpt-5-mini tokenizer. Blue dashed lines mark the 90th percentile for the combined datasets: 32,448 tokens for SecBench.js \& VulcaN and 147,025 tokens for recently crawled \npm packages. Only JavaScript files (with extensions .js, .jsx, .mjs, and .cjs) are included in the count. We exclude TypeScript files because most published \npm packages distribute transpiled JavaScript artifacts for execution, %
  and our analysis focuses on code that is directly executed in production.
  }
  \label{fig:motivation_cdf_token_counts}
\end{figure}

\paragraph{Reasoning beyond handcrafted program models}
As summarized in \autoref{tab:pa_tool_limitations}, many challenges faced by rule-based %
tools are from the need to construct and maintain accurate program models, which is often at odds with scalability and requires substantial manual effort and domain expertise. Moreover, there is a steep increase in the effort required for improving the analysis to cover additional features, once core behaviors have been modeled.
In contrast, learning-based approaches, especially large language models in the current era, provide a promising alternative by utilizing their extensive pre-trained knowledge and reasoning abilities to comprehend complex code patterns, library usages, and dynamic behaviors without requiring exhaustive manual modeling.
As a result, they can adapt more readily to evolving programming practices and software ecosystems, where manually maintaining precise program models becomes increasingly impractical.

\begin{figure*}[t]
  \centering
  \includegraphics[width=0.85\linewidth]{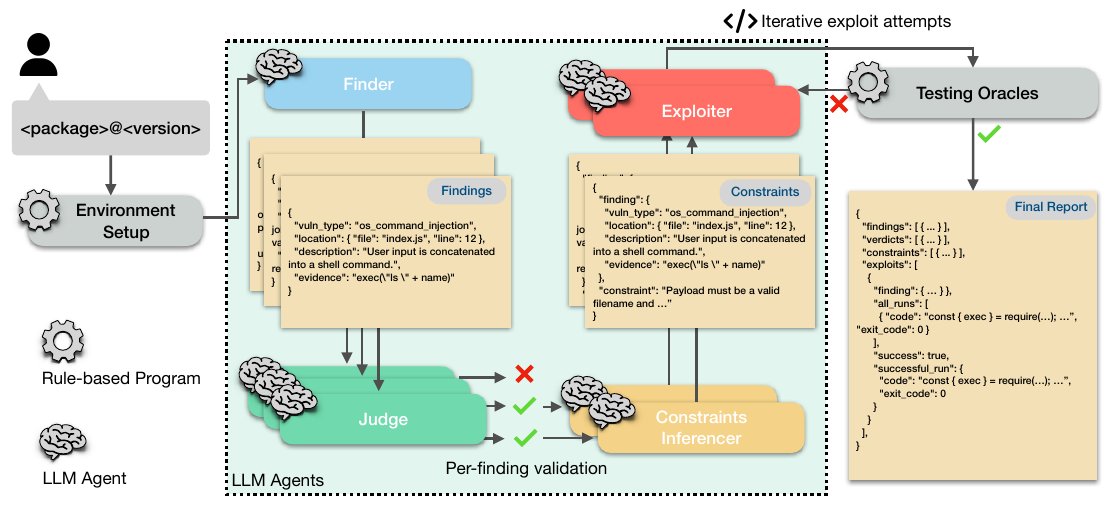}
  \caption{Overview of \toolname.}
  \label{fig:agentic-pipeline}
\end{figure*}

\paragraph{Oracle-Guided Iterative Reasoning}
Another important factor that makes LLM agents well-suited for taint-style vulnerability detection and confirmation tasks in \nodejs packages is that LLM agents can iteratively refine their answers based on feedback from previous unsuccessful attempts. %
In contrast, traditional program-analysis techniques typically perform one-shot reasoning over a fixed program representation. While iterative counter-example guided abstraction refinement (CEGAR) methods~\cite{clarke2003counterexample} have been applied to domains such as model checking, custom algorithm design and significant engineering effort is needed for a specific tool to benefit from CEGAR.  
Moreover, it is
easy to design and implement a testing oracle for taint-style vulnerabilities by observing side effects, commonly used in previous work ~\cite{fast,nodemedic-fine,explode}.
LLM agents can then
generate candidate inputs, execute them against the target package, and refine their reasoning based on observed outcomes.

\paragraph{Favorable code size distribution in the \npm ecosystem}
Despite the strong reasoning capabilities of large language models, current LLMs remain constrained by finite context window sizes, which limit the amount of code that can be processed effectively in a single pass~\cite{chen2021evaluating, fan2023large, wu2023large}.
However, when considering the natural distribution of real-world \npm packages, we observe that in widely used benchmarks and empirical datasets, most \npm packages are sufficiently small that they do not stress the context limits of modern LLMs.
For instance, as shown in \autoref{fig:motivation_cdf_token_counts}, the majority of packages in the VulcaN and SecBench.js datasets contain relatively small amounts of code, with 90th percentile token counts below 32,449. 
Similarly, in the recently crawled \npm packages (which will be discussed in \autoref{subsec:nodemedic-dataset}), the cumulative distribution over token counts exhibits a comparable trend, where 90th percentile token counts are below 147,026.
This measures the total sizes of the codebase, but in reality, an LLM agent will not load the entire codebase into its context window since it can focus on specific files and functions relevant to the vulnerability detection task. However, even considering this, the 90th percentile token counts are still well within the context window sizes of recent LLM models, such as OpenAI's GPT-5 series (400K), Gemini-3 series (1M), and Claude Sonnet 4.5 (200K by default, 1M experimental).

This does not imply that large or complex packages are unimportant; rather, it reflects the natural size distribution of the \npm ecosystem, where most packages are relatively small. As a result, LLM-based approaches are well suited for reasoning about a substantial fraction of real-world packages. Even when packages are larger, an LLM agent can iteratively construct task-relevant context through pattern-based search and package-structure understanding, without requiring dedicated static/dynamic analysis engines for taint/path derivation.

\section{\toolname Design and Implementation}
In this section, we present the design and implementation details of our proposed multi-stage LLM-based agent framework for detecting and confirming vulnerabilities in \nodejs packages.

\subsection{System Overview}

\autoref{fig:agentic-pipeline} illustrates the architecture of \toolname, a multi-stage framework for vulnerability detection and exploit confirmation in \nodejs packages. In practice, end-to-end vulnerability confirmation requires reasoning about candidate locations, constructing executable drivers, and validating exploitability with reliable, automated signals. To make this process tractable and auditable, \toolname decomposes the pipeline into a small number of stages with distinct objectives: the initial finding stage prioritizes high recall, aiming to identify as many potentially vulnerable locations as possible, while the subsequent stages focus on precision by validating exploitability and eliminating false positives.
Accordingly, \toolname organizes analysis as a staged workflow in which candidate findings
are first enumerated, then filtered for exploitability, then augmented with exploitation conditions, and finally validated through execution-based verification. The final stage uses automated execution oracles that determine success based on concrete side effects. 

\subsection{Target Resolution and Execution Context}

The pipeline accepts either a local project path or an \npm identifier in \texttt{package@version} format and operates on a fixed snapshot of the target package. 
We support four taint-style vulnerability classes that are commonly supported by prior program analysis tools, including command injection, code injection, path traversal, and prototype pollution. 
Each class is registered with (i) a natural language vulnerability specification, (ii) goal-oriented exploitation criteria, and (iii) a class-specific execution oracle.
Success is determined by vulnerability-class–specific side effects. These uniform success predicates enable automated validation across heterogeneous vulnerability types.

\subsection{Multi-Stage Vulnerability Reasoning Pipeline}
\paragraph{Candidate Enumeration (Finder)}
The Finder stage performs hypothesis generation by enumerating candidate vulnerabilities through lightweight codebase exploration, including directory traversal, pattern-based search, and source inspection. 
Each candidate is summarized as a structured hypothesis consisting of a vulnerability type, precise source location, supporting code evidence, and a set of potentially reachable APIs. 
This stage intentionally favors over-approximation and prioritizes coverage over precision through prompt design, deferring exploitability assessment and confirmation to subsequent refinement stages. Each candidate is then processed independently through the remainder of the pipeline.

\paragraph{Exploitability Filtering (Judge)}

The Judge stage filters infeasible hypotheses through focused code inspection and lightweight data-flow reasoning, primarily as a reachability check without solving path constraints. Exported APIs are conservatively treated as externally reachable to avoid prematurely discarding viable attack surfaces.
For each candidate, the stage produces a structured verdict consisting of a binary exploitability label and a concise justification.
Only candidate findings deemed potentially exploitable proceed to constraint inference, thereby eliminating false positives early and reducing unnecessary exploration in later stages.

\paragraph{Constraint Inference (Constraints Inferencer)}

Given a validated hypothesis, the Constraints Inferencer stage derives a compact set of actionable exploitation conditions, including likely entry points, required parameters, payload structure, and relevant bypass considerations. 
These constraints summarize the minimal conditions necessary to propagate attacker-controlled input to the vulnerable sink and serve as an explicit interface between exploitability reasoning and exploit synthesis.

By representing exploitation conditions explicitly as structured constraints, this stage provides a clear, structured interface for subsequent exploit synthesis.

\paragraph{Execution-Coupled Exploit Synthesis (Exploiter)}

The Exploiter stage performs execution-coupled synthesis by generating exploits that instantiate the inferred constraints and executing them within the target package environment. 
Each attempt imports the vulnerable module, executes the payload under \nodejs, and records structured results together with full stdout and stderr traces.
Exploit attempts are bounded by iteration limits, and all failed attempts are retained for post hoc analysis. Successful executions are immediately validated by the class-specific oracle (discussed in \autoref{subsec:oracle}) to provide confirmation of exploitability. %

\subsection{Oracle-Guided Execution and Automatic Probing}
\label{subsec:oracle}

A central design choice in \toolname is the use of oracle-guided exploit confirmation to avoid reliance on model self-reporting and manual inspection. 
For each vulnerability class, we define an execution oracle that evaluates concrete side effects produced by exploit attempts.
The execution harness augments each payload with vulnerability-specific probing logic. 
For example, code injection exploits must trigger a predefined marker function; prototype pollution exploits are validated by probing polluted object properties; and path traversal exploits must read and print a prepared sentinel file. 
For OS command injection, sentinel artifacts are removed before each attempt to prevent contamination across runs.

These class-specific probes provide uniform, automated success predicates and enable execution-driven refinement of exploit hypotheses without explicit symbolic constraint solving.

\subsection{Tooling and Coordination Infrastructure}

We organize the supporting toolset into three categories aligned with the stages of vulnerability reasoning. 
First, exploration and inspection tools enable rapid understanding of package structure and relevant logic through directory navigation, source reading, and pattern-based search. These tools are shared across all reasoning stages.
Second, execution and environment interaction tools are restricted to the exploit synthesis stage and provide controlled \nodejs and shell execution, vulnerability-specific harness integration, auxiliary side-effect checks, and optional background process management for long-running services such as web server applications.
Finally, structured reporting and coordination utilities enforce typed submissions for hypotheses, verdicts, constraints, and exploit results, enabling stage-level auditing, failure attribution, and systematic analysis of intermediate reasoning behavior.

\subsection{Implementation}
Each conceptual stage described above is implemented as a dedicated LLM agent. Each agent operates with an isolated context and is responsible for a specific task within the pipeline.

\paragraph{LLM Model Selection}
Since our evaluation involves vulnerability detection and confirmation on unrevealed or unpublished vulnerabilities, we use APIs that have a non-training policy to minimize the risk of data leakage and the possibility of further training LLMs on unrevealed vulnerabilities. To balance performance and cost, we select OpenAI's GPT-5-mini model (\texttt{gpt-5-mini-2025-08-07}) as our LLM backbone. The model is accessed via OpenAI's API platform.

\paragraph{Agent Framework}
We implement our multi-stage agent framework based on LangChain\footnote{https://python.langchain.com/en/latest/index.html}, a popular framework for developing LLM-powered applications. LangChain provides modular components for building complex agent workflows.
 We use a recursion limit of 54 for the agents to prevent infinite or excessive loops during tool usage. 
Considering the non-determinism of \toolname, we allow up to three attempts when no successful exploit is generated. Here, success is defined as observable side effects detected by automated oracles, but not by manual verification.

We adopt a custom multi-stage architecture rather than existing open-source frameworks such as OpenHands~\cite{wang2024openhands} and SWE-agent~\cite{yang2024swe}, to provide explicit stage separation, enabling fine-grained auditing, debugging, and the integration of domain-specific verification components. While these frameworks are effective for general software engineering tasks, they lack native support for stage-level logging and interfaces for vulnerability validation, exploit execution, and oracle-based verification required for our analysis. 
\toolname enables our %
study of LLM agent behavior 
for \npm vulnerability detection and exploit generation 
in a transparent and controllable environment.

\paragraph{Prompt Design} 
We create custom prompts for each agent stage to guide the LLM's reasoning and tool usage. The prompts include clear instructions, the exploitation goal, and an output structure formatted in JSON that contains the expected information for each stage. We refine the prompts iteratively based on initial experiments to improve agent performance, while ensuring no information leakage related to the evaluated packages. Full prompt templates are provided in Appendix~\ref{appendix:prompts}.

\section{Evaluation}

\begin{table*}[t]
  \centering
  \footnotesize
  \caption{Overview of the VulcaN, SecBench.js, NodeMedic, and the newly crawled Wild datasets. ``Raw'' refers to the initial number of vulnerable instances per dataset as used in previous work (VulcaN and SecBench.js) or as received (NodeMedic and Wild). ``Valid'' denotes packages still available on the \npm registry at collection time. ``Used'' is the sampled subset.
   ``Total'' is the count of valid packages and sampled packages. ``Dist. (\%)'' indicates the distribution of each vulnerability type across the combined datasets.}
  \label{tab:dataset_overall}

  \resizebox{\textwidth}{!}{%
  \begin{tabular}{@{} l l r r r r r r r r r r @{}}
    \toprule
    \multirow{2}{*}{\textbf{Vulnerability Type}} &
    \multirow{2}{*}{\textbf{CWE}} &
    \multicolumn{2}{c}{\textbf{VulcaN}} &
    \multicolumn{2}{c}{\textbf{SecBench.js}} &
    \multicolumn{2}{c}{\textbf{NodeMedic}} &
    \multicolumn{2}{c}{\textbf{Wild}} &
    \multirow{2}{*}{\textbf{Total}} &
    \multirow{2}{*}{\textbf{Dist. (\%)}} \\

    \cmidrule(lr){3-4}
    \cmidrule(lr){5-6}
    \cmidrule(lr){7-8}
    \cmidrule(lr){9-10}

    & & \textbf{Raw} & \textbf{Valid}
      & \textbf{Raw} & \textbf{Valid}
      & \textbf{Raw} & \textbf{Used}
      & \textbf{Raw} & \textbf{Used}
      & & \\

    \midrule
    Path Traversal      & CWE-22   & 5  & 3   & 161 & 156 & 0    & 0   & 91 & 65 & 224 & 21.62\% \\
    Command Injection   & CWE-78   & 66 & 58  & 82  & 78  & 1,022 & 130 & 1,004 & 65 & 331 & 31.95\% \\
    Code Injection      & CWE-94   & 22 & 21  & 21  & 21  & 228  & 129 & 452 & 65 & 236 & 22.78\% \\
    Prototype Pollution & CWE-1321 & 67 & 62  & 120 & 118 & 0    & 0   & 104 & 65 & 245 & 23.65\% \\

    \midrule
    \textbf{Total} & &
      160 & 144 &
      384 & 373 &
      1,250 & 259 &
      1,651 & 260 &
      1,036 & 100.00\% \\
    \bottomrule
  \end{tabular}%
  }
\end{table*}

We evaluate \toolname on a variety of datasets to answer the following research questions:
\begin{itemize}[leftmargin=*,itemsep=2pt,topsep=2pt]
  \item \textbf{RQ1}: How effective is \toolname in detecting and confirming vulnerabilities in \nodejs packages?
  \item \textbf{RQ2}: What is the cost of using \toolname?
  \item \textbf{RQ3}: What are the limitations and failure modes of \toolname, and how can they inform future improvements?
\end{itemize}

\subsection{Experiment Setup}
\subsubsection{Datasets}

Our evaluation utilizes four types of datasets: 1) public benchmarks commonly used in vulnerability detection research, 2) a private dataset of real-world \nodejs packages with known vulnerable data paths but no associated CVEs or public exploits, 3) a transformed dataset derived from public benchmarks to assess generalizability, and 4) a crawled dataset of recently released \nodejs packages from the \npm registry to further evaluate performance on unseen data.

\RNparagraph{Public benchmarks}
Following common practice, we use two widely recognized public benchmarks: VulcaN~\cite{vulcan} and SecBench.js~\cite{secbench}. These datasets collected vulnerable \nodejs packages from the \npm registry based on reports from GitHub Advisory, Snyk, Huntr.dev, and the CVE database. For a fair comparison with past work, we select four vulnerability types that have been studied previously: path traversal (CWE-22), OS command injection (CWE-78), code injection (CWE-94), and prototype pollution (CWE-1321). 
We use the same set of packages evaluated in the most recent prior work (\explodejs~\cite{explode}), excluding packages that are no longer available on the \npm registry.

\RNparagraph{Private dataset}
\label{subsec:nodemedic-dataset}
We obtained access to the \nodemedic private dataset~\cite{nodemedic-fine}, which contains real-world \nodejs packages with vulnerable data paths but without associated CVEs or public exploits. 
Although all packages contain vulnerable code paths, no CVEs or other security advisories have been assigned for reasons like low download count or the developers' assumption that the parent package should sanitize the input before calling the vulnerable API. 
This dataset can only be used to evaluate our framework on 
code injection and OS command injection vulnerabilities, the only types supported by \nodemedic.
Considering cost and time constraints, we randomly sample 129 packages with code-injection vulnerabilities and 130 with command-injection vulnerabilities for evaluation. These sample sizes were chosen based on the average number of vulnerable instances per vulnerability type in the VulcaN and SecBench.js datasets after filtering.
This dataset is especially useful for assessing our framework's capability to identify vulnerabilities that are not publicly documented, thus creating a more realistic scenario for vulnerability detection, particularly regarding the issues of LLMs memorizing codes and exploits instead of engaging in genuine reasoning. In this work, we refer to this dataset as NodeMedic.

\RNparagraph{Transformed dataset}
Since the public benchmarks used in this work were released before the knowledge cutoff date of the LLM we chose (GPT-5-mini: May 31, 2024), there is a risk that memorization in the LLM affects the performance and thus it is crucial to evaluate how well our framework generalizes to unseen data. To achieve this, we create a transformed dataset by selecting up to 20 vulnerable instances per CWE from each public dataset, and applying code transformations such as renaming variables and functions, removing comments, and changing formatting. Package names, versions, and links in manifest files are also anonymized. The aim is to generate code that maintains the original semantics and vulnerabilities while being sufficiently different from the LLMs' training data, thus reducing the chances of memorization.
For transforming JavaScript code, we use \texttt{terser}~\cite{terserTerser}, a toolkit for mangling and compressing JavaScript.

\RNparagraph{Crawled dataset}
\label{subsubsec:crawled_dataset}
To further evaluate our framework on unseen data, we crawled recently released \nodejs packages from the \npm registry that were published in December 2025. We consider only newly released or updated packages and collected 17,151 packages during this period. We designed regular expressions to identify potential vulnerable code patterns.
The full regex set is provided in Appendix~\ref{appendix:regex}.
For example, we use \verb@\b(?:eval|Function)\s*\(@ to identify potential code-injection vulnerabilities.
Considering cost and time constraints, we randomly sample 65 packages per vulnerability type that were flagged by the regex patterns. To ensure diversity, for each vulnerability class we discretize three structural metrics (code size, dependency count, and number of files) into coarse buckets and stratify packages by the resulting bucket combinations. When applicable, we additionally ensure that both minified and non-minified artifacts are represented in the sample.
The detailed stratified sampling procedure is described in Appendix~\ref{appendix:sampling}.

\autoref{tab:dataset_overall} provides an overview of the datasets used in our evaluation, including vulnerability counts from VulcaN, SecBench.js, and NodeMedic.

\begin{table*}[t]
  \centering
  \caption{Performance comparison of \toolname against state-of-the-art tools on standard benchmarks.
  ``NM-FINE'' = \nodemedic.
  ``Det.'' = detected, ``Expl.'' = exploited, ``Val.'' = valid.
  \explodejs is evaluated in two modes: ``File'' and ``Pkg''.
The total number of packages and the number of exploits for each tool, both by dataset and overall, are in \textbf{bold} for easier comparison.}
  \label{tab:results_secbench_vulcan}

  \begin{adjustbox}{max width=\textwidth}
  \begin{tabular}{@{} l l r r r r r r r r r r r r @{}}
    \toprule
    \multirow{3}{*}{\textbf{Dataset}} &
    \multirow{3}{*}{\textbf{Vulnerability Type}} &
    \multirow{3}{*}{\textbf{Total}} &
    \multicolumn{2}{c}{\textbf{\fasttool}} &
    \multicolumn{2}{c}{\textbf{NM-FINE}} &
    \multicolumn{4}{c}{\textbf{\explodejs}} &
    \multicolumn{3}{c}{\textbf{\toolname}} \\
    
    \cmidrule(lr){4-5}
    \cmidrule(lr){6-7}
    \cmidrule(lr){8-11}
    \cmidrule(lr){12-14}

    & & &
    \multicolumn{2}{c}{} &
    \multicolumn{2}{c}{} &
    \multicolumn{2}{c}{\textbf{File}} &
    \multicolumn{2}{c}{\textbf{Pkg}} &
    \multicolumn{3}{c}{} \\
    \cmidrule(lr){8-9}\cmidrule(lr){10-11}

    & & &
    \textbf{Det.} & \textbf{Expl.} &
    \textbf{Det.} & \textbf{Expl.} &
    \textbf{Det.} & \textbf{Expl.} &
    \textbf{Det.} & \textbf{Expl.} &
    \textbf{Det.} & \textbf{Expl.} & \textbf{Val.} \\
    \midrule

    \multirow{5}{*}{\textbf{SecBench.js}} &
Path Traversal       & 156 & 105 &  6 &   - &   - &  88 & 79 & 51 & 49 & 155 & 155 & 149 \\
& Command Injection    &  78 &  65 & 60 &  37 &  31 &  56 & 41 &  1 &  1 &  77 &  77 &  76 \\
& Code Injection       &  21 &   8 &  2 &   5 &   1 &   7 &  4 &  3 &  1 &  20 &  20 &  18 \\
& Prototype Pollution  & 118 &   0 &  0 &   - &   - &  53 & 48 &  4 &  2 & 113 & 113 &  89 \\
    \cmidrule(lr){2-14}& \textbf{Total}      & \textbf{373} & 178 & \textbf{68} & 42 & \textbf{32} & 204 & \textbf{172} & 59 & \textbf{53} & 365 & 365 & \textbf{332} \\
    
    \midrule

    \multirow{5}{*}{\textbf{VulcaN}} &
Path Traversal       &   3 &   1 &  0 &   - &   - &   2 &  1 &  1 &  1 &   3 &   3 &   3 \\
& Command Injection    &  58 &  46 & 38 &  15 &   9 &  31 & 16 & 10 &  3 &  54 &  53 &  48 \\
& Code Injection       &  21 &  13 &  5 &   4 &   1 &  10 &  3 &  3 &  0 &  14 &  14 &  12 \\
& Prototype Pollution  &  62 &   0 &  0 &   - &   - &  33 & 31 &  6 &  4 &  58 &  55 &  38 \\
    \cmidrule(lr){2-14}& \textbf{Total}      & \textbf{144} &  60 & \textbf{43} & 19 & \textbf{10} &  76 &  \textbf{51} & 20 &  \textbf{8} & 129 & 125 & 101 \\
    \midrule
    \multicolumn{2}{l}{\textbf{Overall Total}} &
    \textbf{517} &
    238 & \textbf{111} &
    61 & \textbf{42} &
    280 & \textbf{223} &
    79 & \textbf{61} &
    494 & 490 & \textbf{433} \\
    
    \bottomrule
  \end{tabular}
  \end{adjustbox}
\end{table*}

\subsubsection{Baseline tools}

We compare \toolname with three state-of-the-art program-analysis tools for \nodejs package vulnerability detection: 1) \nodemedic~\cite{nodemedic-fine}, 2) \explodejs~\cite{explode}, and 3) \fasttool~\cite{fast}. 
Since our work is the first to utilize LLMs for the complete pipeline of vulnerability detection and confirmation (including PoC generation) in \nodejs packages, we include one LLM-based baseline that is not directly comparable to our method: PoCGen~\cite{pocgen}. PoCGen does not perform vulnerability detection; it generates PoCs based on existing CVE reports by integrating program analysis techniques with LLM reasoning.
We aim to assess whether this combination is necessary or if an LLM-centric, tool-augmented workflow can effectively detect and confirm vulnerabilities.
We configure PoCGen to use the same backend LLM model (GPT-5-mini) as \toolname to ensure a fair comparison. In addition, we run PoCGen on each package up to three times upon failure, matching the maximum number of attempts allowed for \toolname.

\subsubsection{PoC Validation}

Even though both \toolname and the baseline method PoCGen include mechanisms to validate generated PoCs and eliminate false positives, situations may still arise in which the generated PoCs are invalid. In particular, LLM-generated PoCs may trigger the desired side effects while failing to truly exploit the intended vulnerability. To address this issue, we apply an additional manual validation step to all PoCs generated by both \toolname and PoCGen in order to ensure their accuracy. Concretely, among the PoCs that pass automated validation, we further filter out those that exhibit the following behaviors (which we mark as false positives or FPs):
\begin{description}[style=sameline,leftmargin=0.8cm]
  \item[FP1] The PoC introduces a new vulnerability to the runtime environment instead of exploiting an existing one, such as redefining a built-in function to create specific side effects. We mainly observe this behavior in packages with vulnerabilities that exist only on certain operating systems or relies on certain dependencies, and the LLM attempts to emulate such environments by replacing key built-in functions with stubs.
  \item[FP2] The PoC assumes that certain files with specific names exist, where the file names either match the payload or contain the payload. Or, the PoC modifies the environment variables. This is too strong of an assumption to make in practice. This occurs when the package checks for the existence of certain files to decide whether to execute specific code paths.
  \item[FP3] The PoC does not use any public APIs of the package and instead relies on internal code paths, dependencies, test code, or example scripts.
  \item[FP4] For prototype pollution vulnerabilities, the PoC directly uses \texttt{Object} or \texttt{Object.prototype} as part of the arguments passed to the vulnerable package APIs, which is unrealistic in practice because an attacker typically cannot supply these built-in objects as inputs.
  \item[FP5] External tools like web browsers are necessary to %
  trigger the vulnerability. The PoC may simulate this process by directly calling internal functions that these external tools would typically invoke. For a rigorous evaluation of our framework, we classify such PoCs as invalid.
\end{description}

\subsection{RQ1: Effectiveness}
\subsubsection{Comparison with Traditional Program Analysis Tools}

\begin{figure*}[ht]
\centering
\begin{subfigure}[b]{0.2\textwidth}
  \centering
  \includegraphics[width=\textwidth]{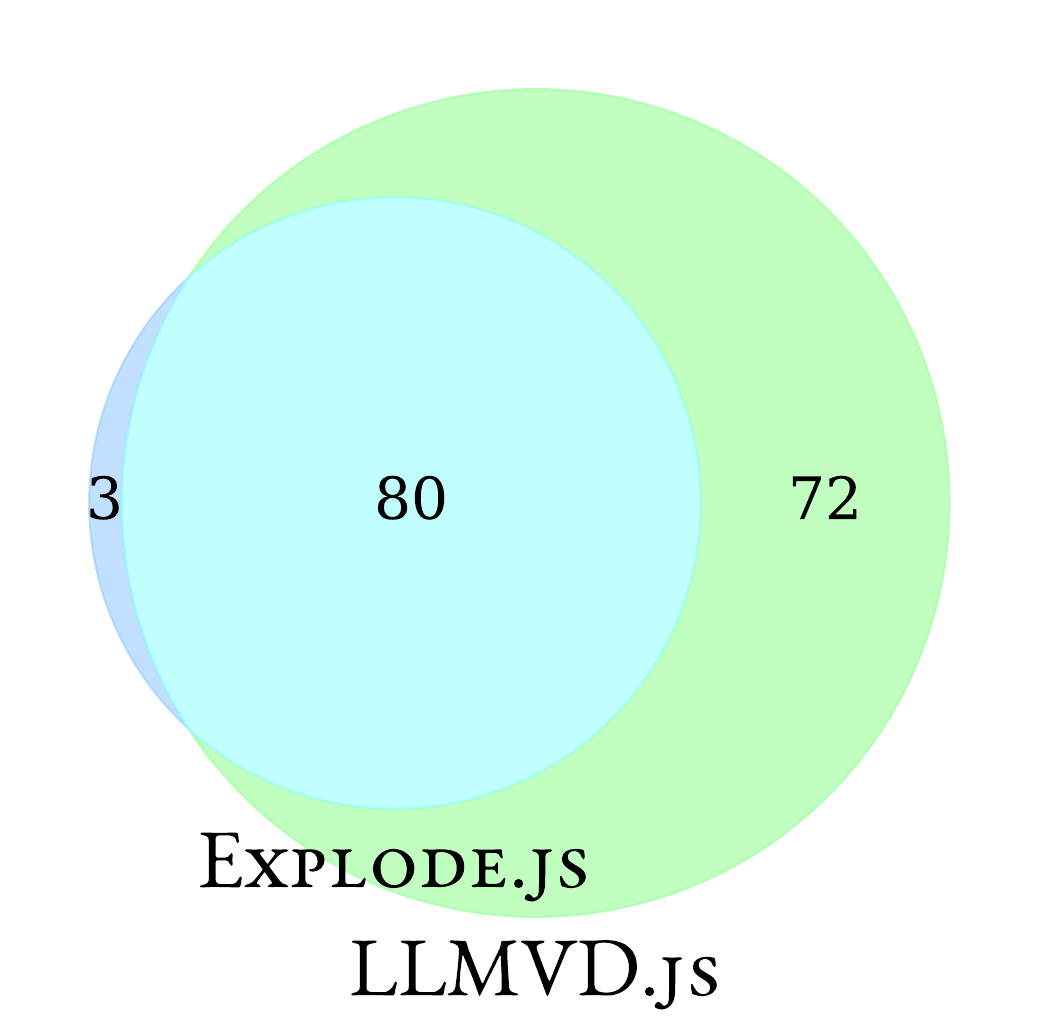}
  \caption{Path Traversal}
  \label{fig:venn_path_traversal}
\end{subfigure}\hfill
\begin{subfigure}[b]{0.28\textwidth}
  \centering
  \includegraphics[width=\textwidth]{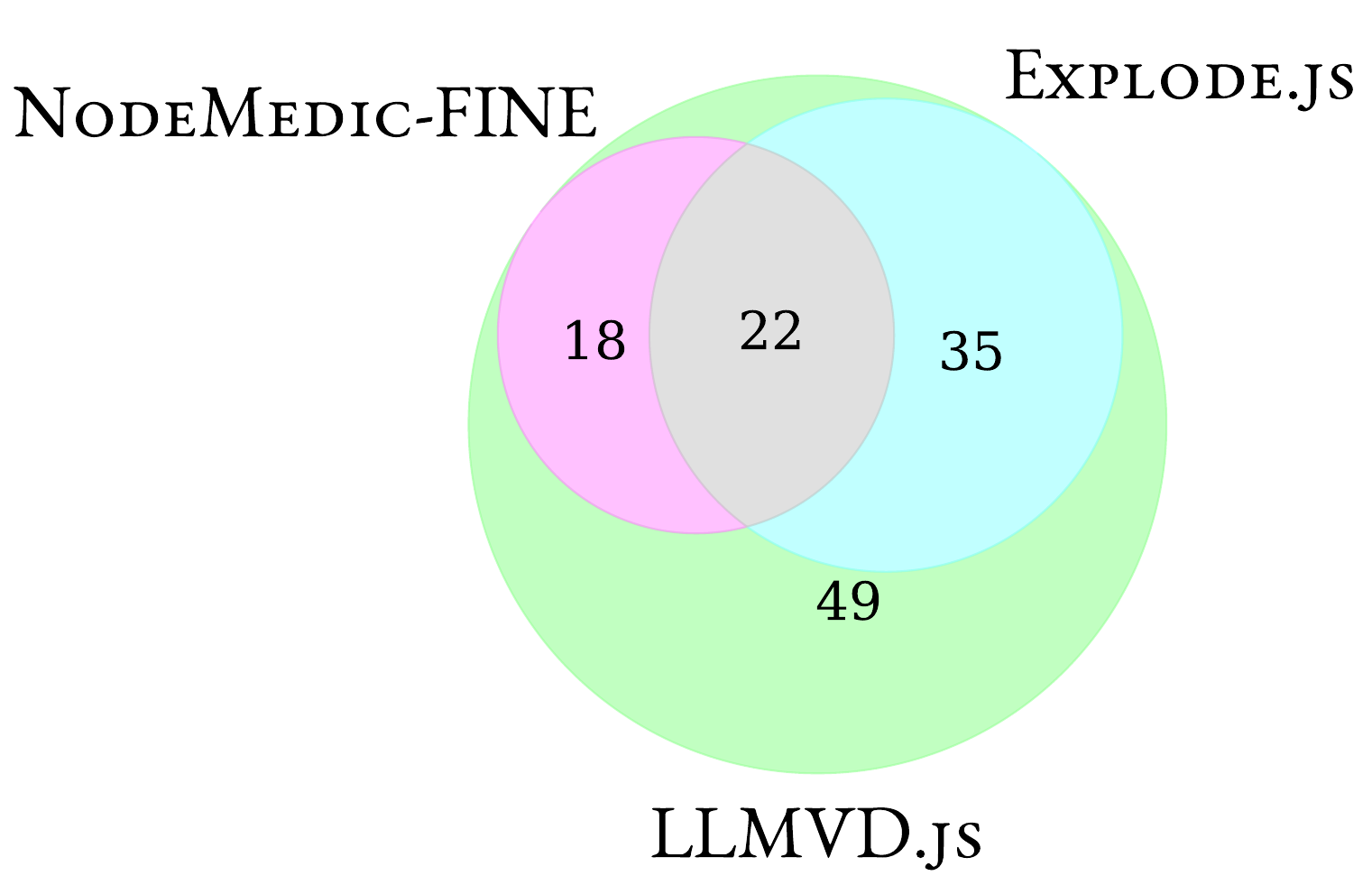}
  \caption{Command Injection}
  \label{fig:venn_os_command_injection}
\end{subfigure}\hfill
\begin{subfigure}[b]{0.28\textwidth}
  \centering
  \includegraphics[width=\textwidth]{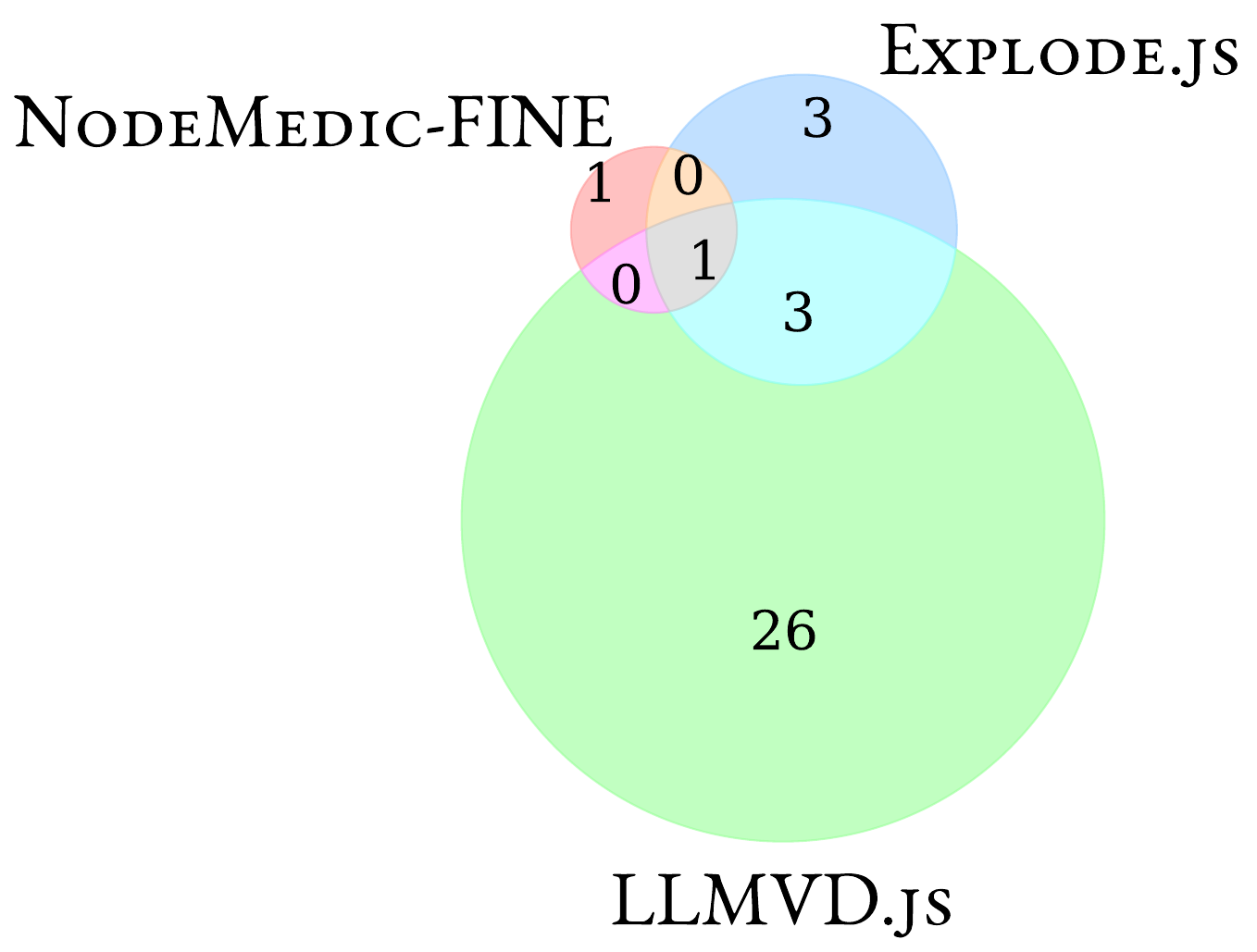}
  \caption{Code Injection}
  \label{fig:venn_code_injection}
\end{subfigure}\hfill
\begin{subfigure}[b]{0.2\textwidth}
  \centering
  \includegraphics[width=\textwidth]{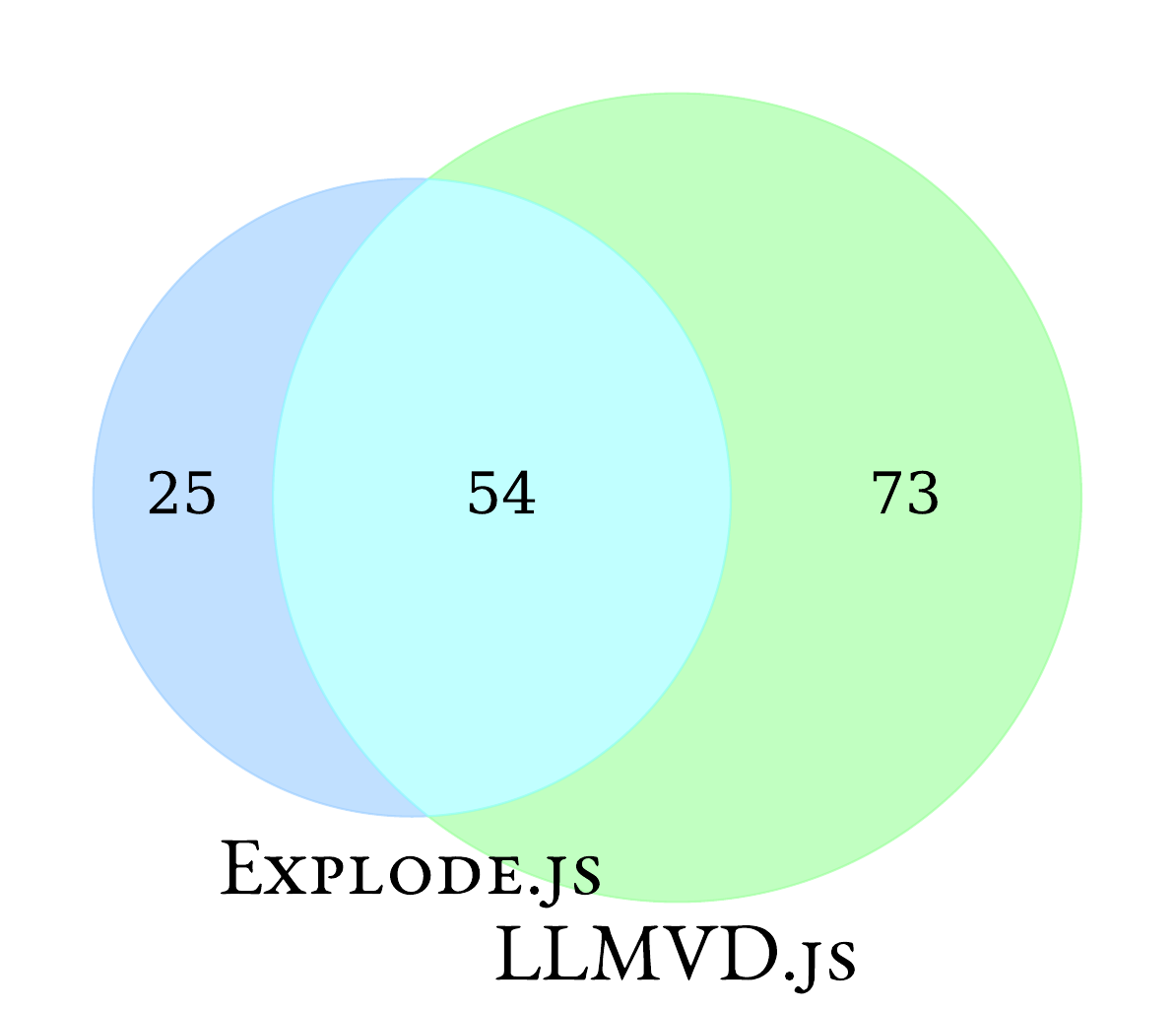}
  \caption{Prototype Pollution}
  \label{fig:venn_prototype_pollution}
\end{subfigure}
\vspace{-5pt}
\caption{Venn diagram overlaps for vulnerability types.}
\label{fig:venn_four_horizontal}
\end{figure*}

\begin{table*}[t]
  \centering
  \caption{Performance comparison of \toolname against PoCGen on 299 overlapping SecBench.js packages.
  ``Expl.'' = exploited, ``Val.'' = valid exploit, ``Avg Cost'' = average LLM API cost per package.
  }
  \label{tab:pocgen_comparison}

  \begin{tabular}{@{} l r r r r r r r @{}}
    \toprule
    \multirow{2}{*}{\textbf{Vulnerability Type}} &
    \multirow{2}{*}{\textbf{Total}} &
    \multicolumn{3}{c}{\textbf{PoCGen}} &
    \multicolumn{3}{c}{\textbf{\toolname}} \\

    \cmidrule(lr){3-5}
    \cmidrule(lr){6-8}

    & & \textbf{Expl.} & \textbf{Val.} & \textbf{Avg Cost}
    & \textbf{Expl.} & \textbf{Val.} & \textbf{Avg Cost} \\
    \midrule

    Path Traversal       & 117 & 112 & 108 & \$0.089 & 116 & 112 & \$0.050 \\
    Command Injection    &  67 &  63 &  62 & \$0.124 &  67 &  66 & \$0.068 \\
    Code Injection       &  12 &  10 &   9 & \$0.172 &  12 &  11 & \$0.099 \\
    Prototype Pollution  & 103 &  93 &  74 & \$0.079 &  99 &  79 & \$0.135 \\
    \midrule
    \textbf{Total} & 299 & 
      278 & 
      253 & 
      \$0.097 & 
      294 & 
      268 & 
      \$0.085 \\
    
    \bottomrule
  \end{tabular}
\end{table*}

\autoref{tab:results_secbench_vulcan} shows the performance comparison of \toolname against three leading program-analysis tools (\nodemedic, \fasttool, and \explodejs) on the SecBench.js and VulcaN datasets. It's important to note that \explodejs operates in two modes: "File" mode, where the tool analyzes a specific file, and "Pkg" mode, where the tool examines the entire package without prior knowledge of which file contains the vulnerability. We include both modes in our comparison because "File" mode is used in the original \explodejs paper~\cite{explode}, while "Pkg" mode is more realistic for comprehensive vulnerability detection.

\toolname demonstrates significant advantages over all baselines in both datasets across all types of vulnerabilities. Notably, \toolname successfully detects and generates valid exploits for 433 out of 517 vulnerable packages in both datasets, achieving an overall confirmation rate of 83.75\%. In contrast, the best-performing baseline, which is not realistic in an end-to-end detection and confirmation setting, \explodejs in "File" mode, manages to confirm only 223 vulnerabilities, resulting in a confirmation rate of 43.13\%.

Examining the distribution of successful exploit generations by various vulnerability types, \autoref{fig:venn_four_horizontal} shows that while \toolname significantly outperforms the baselines, there are still packages where the baselines succeed while \toolname does not. This suggests that rule-based tools can offer complementary strengths in specific situations, which we discuss further in \autoref{subsec:eval_limit}.

\subsubsection{Comparison with PoCGen}

As a comparison with a prior LLM–program analysis hybrid method, we investigate whether a carefully designed pipeline with program-analysis components can still improve LLM reasoning in terms of both performance and cost, given the rapid advancement of LLMs.
This comparison with PoCGen~\cite{pocgen} is %
favorable to PoCGen. PoCGen is provided with a CVE report when generating exploits, whereas \toolname operates as an end-to-end framework that detects and confirms vulnerabilities without being provided any prior knowledge of the target packages beyond their source code. Nevertheless, PoCGen is the closest available LLM-based work that targets a partially overlapping problem setting.

For a fair comparison, we only evaluate the overlapping packages from the PoCGen~\cite{pocgen} and our SecBench.js~\cite{secbench} datasets. 
We also removed any packages whose vulnerability reports were deleted at the time of our evaluation, as PoCGen cannot initiate the pipeline without the report. This results in a total of 299 packages. %
\autoref{tab:pocgen_comparison} presents the comparison results. \toolname outperforms PoCGen in terms of valid exploit generation across all vulnerability types 
and incurs lower LLM API costs in three of the four vulnerability types and achieves lower overall cost.

We investigate why the LLM-program analysis hybrid design in PoCGen does not result in better performance or, at the very least, lower costs. We summarize our observations in the following points:
(1) CodeQL AST/locations weren't converted to compact facts, so the LLM still tried to resolve references itself.
(2) LLMs can infer taint flows and definitions directly from raw code, so verbose CodeQL snippets offered little additional, non-redundant signal.
(3) when a refinement fails, the refiner generates multiple slightly different prompt variants and sends each to the model, so one failure becomes many near‑duplicate LLM requests, increasing API calls and token usage;
(4) single prompts often embed overlapping sections (examples, descriptions, snippets) that inflate tokens per call beyond a concise agent prompt.
These emphasize the need to better design program analysis components that can effectively assist LLM in reasoning for this task. We discuss potential future directions based on the failure modes of \toolname that we observed in \autoref{subsec:eval_limit}.

\subsubsection{Transformed dataset}

\begin{table*}[t]
  \centering
  \caption{Performance comparison of \toolname against state-of-the-art tools on recently released \nodejs packages from the \npm registry and the private NodeMedic dataset.
  ``NM-FINE'' = \nodemedic,
  ``Det.'' = detected, ``Expl.'' = exploited, ``Val.'' = valid. ``-'' indicates that reporting is not applicable for the NodeMedic dataset.
  The total number of packages and the number of exploits for each tool, both by dataset and overall, are in \textbf{bold} for easier comparison.}
  \label{tab:results_wild}

  \begin{adjustbox}{max width=\textwidth}
  \begin{tabular}{@{} l l r r r r r r r r r r @{}}
    \toprule
    \multirow{2}{*}{\textbf{Dataset}} &
    \multirow{2}{*}{\textbf{Vulnerability Type}} &
    \multirow{2}{*}{\textbf{Total}} &
    \multicolumn{2}{c}{\textbf{\fasttool}} &
    \multicolumn{2}{c}{\textbf{NM-FINE}} &
    \multicolumn{2}{c}{\textbf{\explodejs}} &
    \multicolumn{3}{c}{\textbf{\toolname}} \\

    \cmidrule(lr){4-5}
    \cmidrule(lr){6-7}
    \cmidrule(lr){8-9}
    \cmidrule(lr){10-12}

    & &
    & \textbf{Det.} & \textbf{Expl.}
    & \textbf{Det.} & \textbf{Expl.}
    & \textbf{Det.} & \textbf{Expl.}
    & \textbf{Det.} & \textbf{Expl.} & \textbf{Val.} \\
    \midrule

    \multirow{3}{*}{\textbf{NodeMedic}} &
Command Injection    & 130 & 84 & 78 & 130 & 71 & 10 & 5 & 129 & 128 & 120 \\
    & Code Injection       & 129 & 78 & 29 & 129 & 56 & 22 & 9 & 128 & 128 & 124 \\
    \cmidrule(lr){2-12}
     & \textbf{Total} & \textbf{259} &
       162 & \textbf{107} &
      259 & \textbf{127} &
      32 & \textbf{14} &
      257 & 256 & \textbf{244} \\
    \midrule

    \multirow{5}{*}{\textbf{Wild}} &
    Path Traversal      & 65 & 0 & 0 & - & - & 0 & 0 & 44 & 37 & 6 \\
    & Command Injection   & 65 & 3 & 2 & 0 & 0 & 0 & 0 & 28 & 26 & 17 \\
    & Code Injection      & 65 & 1 & 0 & 0 & 0 & 0 & 0 & 20 & 17 & 12 \\
    & Prototype Pollution & 65 & 0 & 0 & - & - & 0 & 0 & 20 & 4  & 1 \\
    \cmidrule(lr){2-12}
     & \textbf{Total} & \textbf{260} &
      4 & \textbf{2} &
      0 & \textbf{0} &
      0 & \textbf{0} &
      112 & 84 & \textbf{36} \\
    \midrule

    \multicolumn{2}{l}{\textbf{Overall Total}} & \textbf{519} &
       166 & \textbf{109} &
      259 & \textbf{127} &
      32 & \textbf{14} &
      369 & 340 & \textbf{280} \\
    \bottomrule
  \end{tabular}
  \end{adjustbox}
\end{table*}

We compared the number of packages that \toolname could successfully exploit before and after transformation on the two sampled public benchmarks (VulcaN and SecBench.js), counting only exploits with manually verified valid PoCs. Among the 143 sampled packages, \toolname successfully generated valid PoCs for 108 packages on the original (untransformed) datasets. On the transformed dataset, \toolname generated valid PoCs for 107 packages, missing one previously successful package (\texttt{lodash@4.17.15}) and yielding no additional successful exploits.
A detailed case study of this package %
is provided in \autoref{subsec:eval_limit}.
This result indicates that \toolname generalizes well to unseen data that are syntactically different from the LLM training data, demonstrating robustness against potential memorization effects.

\subsubsection{Private NodeMedic dataset}

\autoref{tab:results_wild} shows a comparison of \toolname with state-of-the-art tools on the private NodeMedic dataset~\cite{nodemedic-fine} (\autoref{subsec:nodemedic-dataset}). \toolname generated valid PoCs for 244 out of a total of 259 vulnerable packages, achieving a valid PoC generation rate of 94.2\%. In contrast, \nodemedic produced working PoCs for only 127 (49.0\%). Despite the significant performance gap, \toolname missed 5 vulnerable packages (3 Code Injection and 2 Command Injection) that were successfully exploited by \nodemedic (more in \autoref{subsec:missed-rule-based}). %

\subsubsection{Crawled \npm packages in the wild}

This dataset includes randomly sampled 65 packages per vulnerability type (path traversal, OS command injection, code injection, and prototype pollution) that were flagged by our regex patterns, resulting in a total of 260 packages (\autoref{subsubsec:crawled_dataset}).
Among the rule-based program analysis tools, only \fasttool detected 4 vulnerable packages (3 command injection and 1 code injection) and successfully exploited 2 of them (2 command injection). \nodemedic and \explodejs did not detect any vulnerabilities in these packages.

In contrast, \toolname detected 112 packages that it identified as potentially vulnerable and successfully exploited 84 of them to produce the required side effects. Among the 84 exploited packages, 36 generated proofs of concept (PoCs) that were deemed valid after manual inspection. The detailed results are presented in \autoref{tab:results_wild}.

We analyze the non-validated cases by categorizing them into two groups: (1) environment-dependent exploits that require specific system configurations or dependencies, where the LLM attempts to emulate the environment by introducing proxies (e.g., stubs or redefined built-ins); and (2) executions through internal code paths such as tests or examples that are not part of the public API. Our manual validation adopts a conservative policy that excludes these categories, which likely underestimates the number of true vulnerabilities. We are currently evaluating cases where the exploits were not deemed valid to determine whether they correspond to real vulnerabilities.

These results highlight a limitation of current LLM-based agents: while they can synthesize executable exploits, they often rely on overly permissive assumptions about entry points and execution conditions. Without explicit guidance, the agent may use internal code paths instead of public APIs or emulate missing environments beyond the intended exploit boundary. These issues point to the need for prompt tuning to clearly define the detection boundary, including public APIs and environment and use-case assumptions.

We have reported all 36 validated vulnerabilities to the respective package maintainers and are in the process of responsible disclosure. So far, we have received acknowledgments from 3 maintainers.

\subsubsection{Fine-Grained Analysis of Detection Results}

\begin{table}[H]
\centering
\setlength{\tabcolsep}{6pt}
\caption{Summary of file-level unmatched rates by dataset (all vulnerability types combined) for Finder and Judge stages.}
\label{tab:file-level-unmatched-small}
\begin{tabular}{lccc}
\hline
Metric & SecBench.js & VulcaN & All \\
\hline
Finder Findings Unmatched & 22.94\% & 28.80\% & 24.77\% \\
Judge Findings Unmatched & 20.89\% & 29.15\% & 23.32\% \\
Finder GT Unmatched & 4.29\% & 10.60\% & 6.11\% \\
Judge GT Unmatched & 4.83\% & 14.57\% & 7.63\% \\
\hline
\end{tabular}
\end{table}

\begin{figure*}[ht]
\centering

\begin{subfigure}[t]{0.45\textwidth}
  \begin{minipage}[t][0.35\textheight][t]{\linewidth}
    \vspace{0pt}
    \centering
    \includegraphics[width=\linewidth,height=0.395\textheight,keepaspectratio]{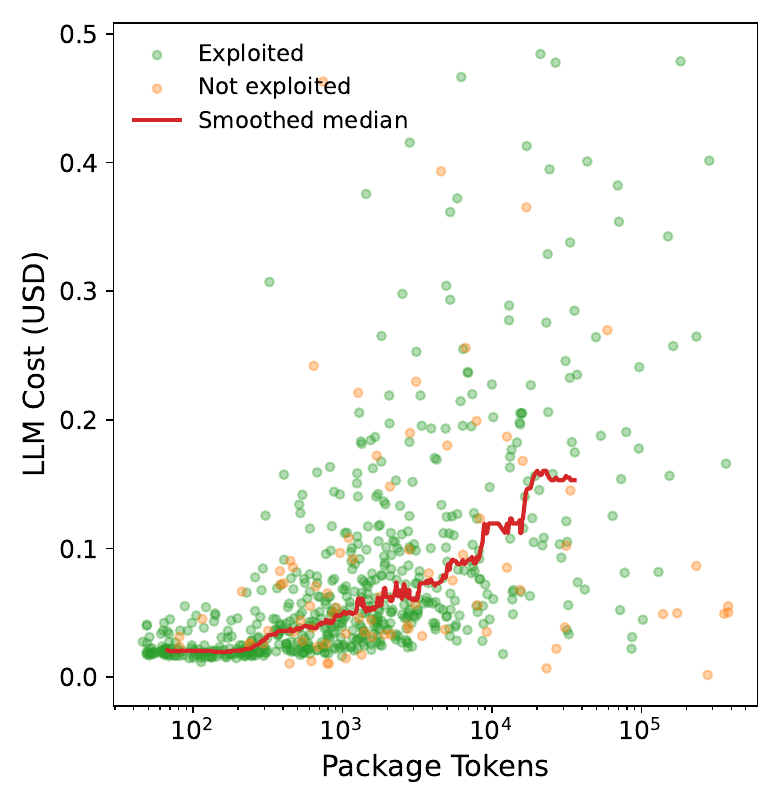}
    \caption{LLM cost vs. package size}
    \label{fig:rq4_cost_vs_tokens}
  \end{minipage}
\end{subfigure}\hfill
\begin{subfigure}[t]{0.53\textwidth}
  \begin{minipage}[t][0.35\textheight][t]{\linewidth}
    \centering

    \begin{subfigure}[t]{\linewidth}
      \begin{minipage}[t][0.17\textheight][t]{\linewidth}
        \vspace{0pt}
        \centering
        \includegraphics[width=0.98\linewidth,height=0.205\textheight,keepaspectratio]{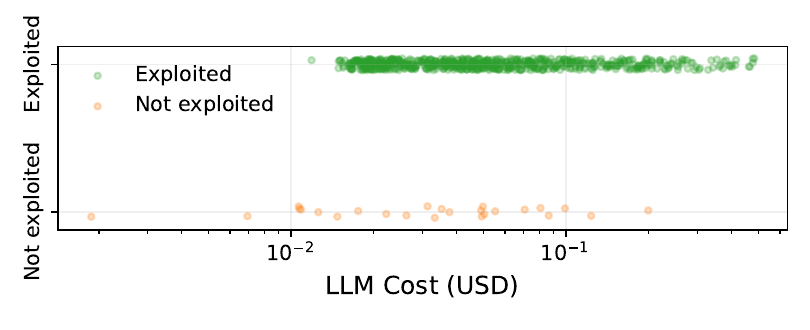}
        \vfill
        \caption{Exploit success vs. LLM cost}
        \label{fig:rq4_success_vs_cost}
      \end{minipage}
    \end{subfigure}

    \vspace{7.7mm}

    \begin{subfigure}[t]{\linewidth}
      \begin{minipage}[t][0.17\textheight][t]{\linewidth}
        \centering
        \includegraphics[width=0.98\linewidth,height=0.205\textheight,keepaspectratio]{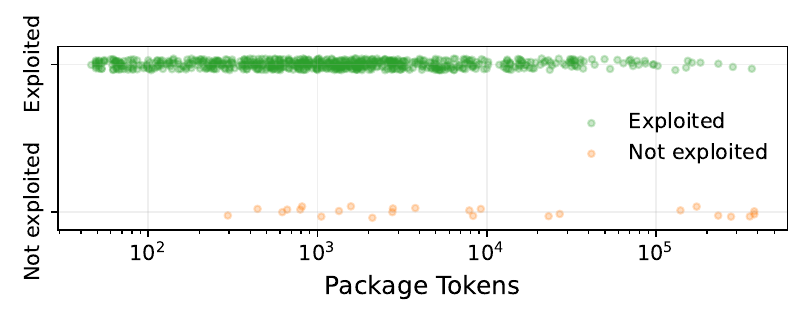}
        \vfill
        \caption{Exploit success vs. package size}
        \label{fig:rq4_success_vs_tokens}
      \end{minipage}
    \end{subfigure}

  \end{minipage}
\end{subfigure}

\vspace{30pt}

\caption{Cost, package size, and exploit success trade-offs. (a) LLM API cost increases with package token count, with the red curve showing the smoothed median. 16 packages with extreme large token counts are omitted for better visualization. (b) Distribution of exploited and non-exploited cases across different LLM cost levels. (c) Distribution of exploited and non-exploited cases across package token counts.}
\label{fig:rq4_cost_token_success}
\end{figure*}

\toolname may report multiple findings per package, and some reports do not match any benchmark-labeled vulnerable file. Because benchmark labels are not guaranteed to be exhaustive, we do not automatically treat unmatched reports as false positives. Instead, we quantify mismatch behavior with file-level coverage metrics.

Here, ``file-level'' means the benchmark ground truth includes the file location for each vulnerability, and \toolname output includes a predicted vulnerable file path for each finding; we count a match only when these file locations agree. At a high level, we measure mismatch from two perspectives at both Finder and Judge stages: (i) report-side mismatch (captured by the Finder/Judge Findings Unmatched rows in \autoref{tab:file-level-unmatched-small}), i.e., what fraction of tool-reported findings cannot be matched to benchmark-labeled vulnerable files, and (ii) ground-truth-side miss rate (captured by the Finder/Judge GT Unmatched rows in \autoref{tab:file-level-unmatched-small}), i.e., what fraction of benchmark vulnerable files are not covered by any reported finding. These metrics are formally defined in Appendix \ref{app:unmatched}. We report these four percentages for each dataset in \autoref{tab:file-level-unmatched-small}. The detailed file-level unmatched results by dataset and vulnerability type are presented in \autoref{tab:file-level-unmatched-summary} in Appendix \ref{app:unmatched}.

Overall, the file-level mismatch rates indicate a favorable tradeoff between over-reporting and missed detections. While \toolname produces a non-trivial fraction of unmatched reports (approximately one quarter), the ground-truth miss rate remains low (around 6–8\%), suggesting that most benchmark-labeled vulnerabilities are successfully covered. We emphasize that unmatched findings arise from two sources: incomplete benchmark labeling and the tool’s over-approximation of potential vulnerabilities.

\subsection{RQ2: Cost}

We compute the average LLM API cost incurred by \toolname across different vulnerability types. The average cost per package is \$0.051 for path traversal, \$0.068 for OS command injection, \$0.107 for code injection, and \$0.136 for prototype pollution, with an overall average cost of \$0.089 per package across all samples. When restricting to successfully exploited packages, \toolname spends an average of \$0.084 per valid exploit. Because \toolname may generate multiple exploit candidates for a single package by targeting different vulnerable locations, we additionally report an amortized cost per valid exploit of \$0.050, computed by dividing the total LLM cost by the number of valid exploits. These results indicate that \toolname achieves effective exploit generation with modest and well-controlled LLM usage costs across vulnerability types.

We further analyze the relationship between LLM cost and exploit success using the per-package distributions. \autoref{fig:rq4_cost_vs_tokens} shows that LLM cost increases with package token count, and that successful exploits are observed across a wide range of token usage levels. \autoref{fig:rq4_success_vs_cost} relates exploit outcomes to the incurred LLM cost and shows no clear monotonic relationship between exploit success and LLM cost, as both exploited and non-exploited packages are distributed 
across similar cost ranges.

\subsection{RQ3: Limitations and Failure Modes}
\label{subsec:eval_limit}
\subsubsection{Impact of Package Size}

\begin{figure*}[t]
\centering

\includegraphics[width=0.98\textwidth]{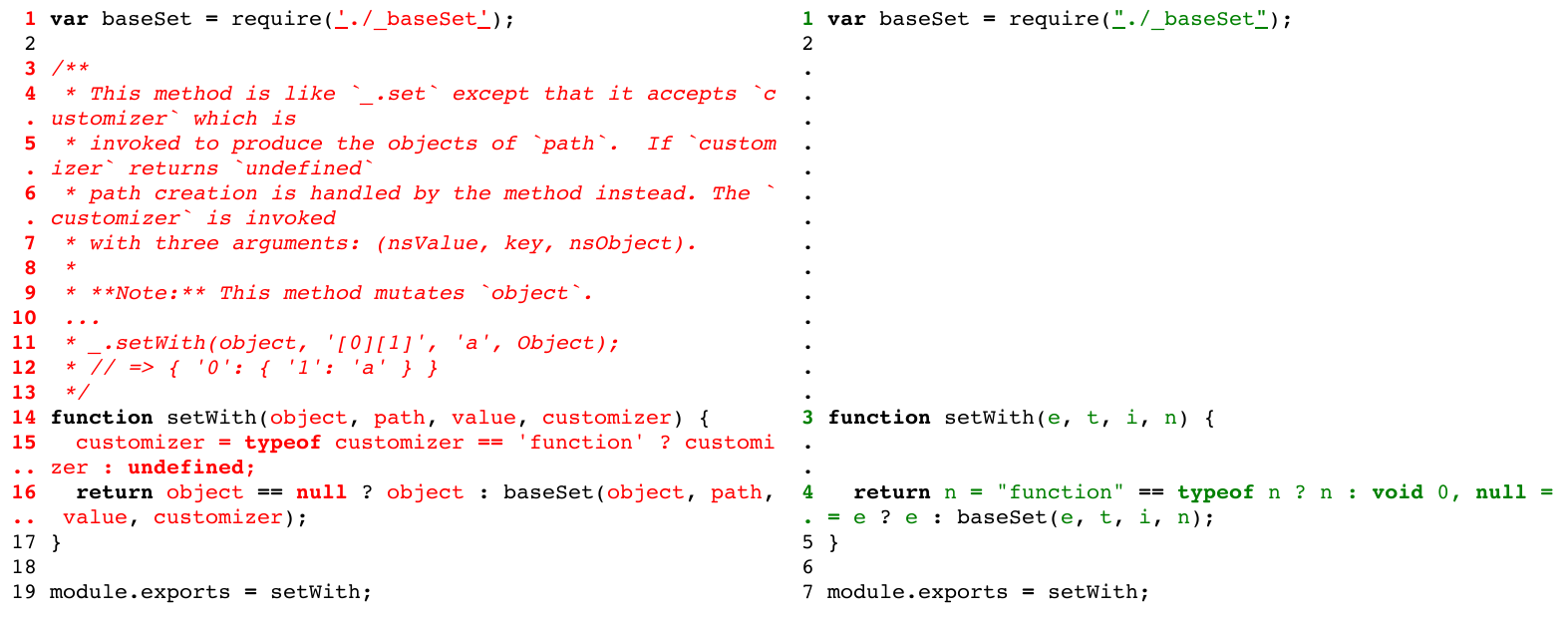}

\caption{One example of the vulnerable sinks in \texttt{lodash@4.17.15} before (left) and after (right) transformation. In ``Before transformation'', some parts of the comments were omitted for brevity and replaced with ``...''.}
\label{fig:lodash_compare}
\end{figure*}

A common thought is that larger packages are harder to exploit because they tend to involve longer and more complex code paths, which can make it more difficult to identify the relevant data flows and construct a working exploit. \autoref{fig:rq4_cost_token_success}(c) examines this relationship by plotting exploit outcomes against package token count. The plot shows that successful exploits are common for small and medium-sized packages, while the proportion of non-exploited cases increases as package size grows.

\subsubsection{A case study of missed generalization}

To better understand the generalization capabilities of \toolname, we investigate the single package that was successfully exploited on the original dataset but missed on the transformed dataset: \texttt{lodash@4.17.15}, which contains prototype-pollution vulnerabilities. In detecting the transformed version, \toolname did not identify any potential vulnerabilities and therefore stopped at the finder stage.
\autoref{fig:lodash_compare} shows a side-by-side comparison of one of the many vulnerable sinks in both the original and transformed code. The transformation included renaming variables and functions, removing comments, and altering the formatting.

In the original package, \toolname reported eight vulnerability findings and successfully identified prototype pollution across multiple relevant files (e.g., \_baseAssignValue.js, \_baseSet.js, and \_baseMerge.js). The agent’s reasoning benefited from semantic cues such as informative variable names and developer comments, which helped it interpret code intent and localize vulnerable behaviors. In contrast, on the transformed package, \toolname produced zero findings and hit the recursion limit (54 iterations). The logs indicate that the agent spent most of its budget attempting to navigate and interpret the obfuscated codebase, becoming effectively lost among approximately 1,046 JS files without meaningful semantic hints. For example, it repeatedly revisited file-tree listings and performed pattern searches, but failed to form a coherent understanding of the code necessary to confirm vulnerabilities.

This case study suggests that certain obfuscation and anonymization patterns can effectively blind LLM-based agents by removing the semantic cues they rely on for navigation and comprehension, leading to substantially degraded vulnerability detection and confirmation. A systematic characterization of which transformations cause these failures, and how to mitigate them, is beyond the scope of this paper, and we leave detailed studies to future work.

\subsubsection{Vulnerabilities Missed by \toolname but Detected by Rule-based Tools} \label{subsec:missed-rule-based}

Most of the vulnerable packages that \toolname overlooked but rule-based program analysis tools successfully exploited are due to manual validations. The LLM believes it has generated a valid PoC and exits, but the PoCs are manually rejected. We discuss this in \autoref{subsubsec:invalid}.
Ruling out this part, there are 8 packages in all of SecBench.js, VulcaN and NodeMedic datasets that \toolname missed but traditional tools successfully exploited. We manually inspected these packages and found that they mainly fall into two categories:
(1) The tool identified the location of the vulnerability but concluded that it had sufficient sanitization or had already been patched, so it did not report it.  
(2) The tool loaded a large but irrelevant portion of the codebase into the context, which caused confusion and led to missing the vulnerability. Although these issues occur in only 4 out of the 776 evaluated packages, they highlight opportunities to further improve the current \toolname pipeline.

\subsubsection{Invalid PoC Generations and False Positives}
\label{subsubsec:invalid}

\begin{table*}[t]
  \centering
  \caption{Analysis of invalid exploit reasons by vulnerability type.
  For each vulnerability type, shows the total number of packages with invalid exploits and breakdown by reason with percentages.}
  \label{tab:invalid_exploits}

  \begin{adjustbox}{max width=\textwidth}
  \begin{tabular}{@{} l r r r r @{}}
    \toprule
    \textbf{Invalid Reason} & \textbf{Path Traversal} & \textbf{Command Injection} & \textbf{Code Injection} & \textbf{Prototype Pollution} \\
    \midrule

    \textbf{Total Packages (Invalid)} & 6 & 14 & 8 & 41 \\
    \midrule

    FP1: Emulated Environment & - & 11 (78.6\%) & 5 (62.5\%) & 4 (9.8\%) \\
    FP2: Strong Assumptions & 1 (16.7\%) & 1 (7.1\%) & - & 1 (2.4\%) \\
    FP3: Non-Public API & 5 (83.3\%) & 1 (7.1\%) & 3 (37.5\%) & 2 (4.9\%) \\
    FP4: Direct Object Use & - & - & - & 34 (82.9\%) \\
    FP5: External Tools Needed & - & 1 (7.1\%) & - & - \\
    
    \bottomrule
  \end{tabular}
  \end{adjustbox}
\end{table*}

\autoref{tab:invalid_exploits} summarizes why some PoCs that pass automated validation are still deemed invalid after manual inspection. Prototype pollution contributes the largest number of invalid cases (41 packages), and the dominant failure mode is FP4, where the PoC unrealistically passes \texttt{Object} or \texttt{Object.prototype} as an input to the target API (34/41, 82.9\%). For OS command injection, most invalid PoCs fall under FP1 (11/14, 78.6\%), where the agent emulates missing environments or dependencies by redefining built-ins or stubbing key functionality, and a small fraction require external tools (FP5). For code injection, invalid cases are split between environment emulation (FP1: 5/8, 62.5\%) and reliance on non-public/internal code paths (FP3: 3/8, 37.5\%). For path traversal, the primary issue is FP3 (5/6, 83.3\%), indicating that the agent often constructs PoCs that exercise internal code paths, tests, or examples rather than public-facing APIs. Overall, these invalid generations concentrate in a few recurring, vulnerability-specific failure modes, which motivates future work on adding rule-based checks or LLM-based guardrails to discourage unrealistic assumptions and improve PoC validity.

Overall, the evaluations do not suggest that LLMs universally outperform classical tools, but instead motivates rethinking the role of program analysis as a complementary technique, particularly in scenarios where formal guarantees or deep semantic reasoning are required.

\section{Threats to Validity}

Even though we tried to design an evaluation to exclude the memorization effect of LLMs as much as possible, there are still some threats to validity that may affect the conclusions drawn from our experiments. First, the code transformations we applied to create the transformed dataset may not be sufficient to completely eliminate the memorization effect, especially for larger models that may have seen similar code snippets during training. Future work could explore more sophisticated transformation techniques or use entirely synthetic datasets to further mitigate this threat. Second, the NodeMedic dataset has distribution bias compared with wild \nodejs packages since they only include those that \nodemedic~\cite{nodemedic-fine} flagged as potentially vulnerable. Therefore, the performance of our framework on this dataset may not fully reflect its effectiveness in real-world scenarios. Third, our evaluation focuses on specific vulnerability types (i.e., path traversal, OS command injection, code injection, and prototype pollution), which may limit the generalizability of our findings to other types of vulnerabilities. Future work could extend the evaluation to a broader range of vulnerability types to assess the versatility of our framework.

\section{Conclusion}

In this work, we demonstrate the strong capabilities of state-of-the-art LLMs in detecting and confirming vulnerabilities in \nodejs packages using LLM-centric, tool-augmented reasoning without dedicated static/dynamic analysis engines for taint/path derivation. As LLMs continue to improve in reasoning and code understanding, the benefits of tightly integrating traditional program analysis techniques, as explored in prior work, may quickly diminish. We believe it is therefore promising to evaluate current LLM capabilities from lower-level perspectives and to carefully design program analysis tools that complement LLMs in ways that cannot be easily achieved through model scaling or architectural improvements alone.

\bibliographystyle{ACM-Reference-Format}
\bibliography{ref}
\clearpage
\section*{Appendix}
\appendix
\section{Regular Expressions for Filtering Crawled \npm Packages}\label{appendix:regex}

Here are the regular expressions we used to filter potentially vulnerable packages from the crawled \npm packages for each vulnerability type:%

\paragraph{Code Injection}
\begin{itemize}[label=--]
  \item \verb@\b(?:eval|Function)\s*\(@: eval/Function usage.
\end{itemize}

\paragraph{Command Injection}
\begin{itemize}[label=--]
  \item \verb@child_process\s*\.\s*(?:exec|spawn)@: child\_process exec/spawn usage.
  \item \verb@require\(\s*['\"]child_process['\"]\s*\)@: require ('child\_process').
  \item \verb@\b(?:execSync|spawnSync)\s*\(@: execSync/spawnSync usage.
\end{itemize}

\paragraph{Path Traversal}
\begin{itemize}[label=--]
  \item \verb@\bpath\s*\.\s*(?:join|resolve)\s*\([^)]*\b(?:req\s*\.\s*(?:params|query|body|headers|url|originalUrl)|process\s*\.\s*env)\b[^)]*\)@: path.join/resolve fed by req.* or env.
  \item \verb@\b(?:fs|node:fs)\s*\.\s*(?:readFile|readFileSync|writeFile|writeFileSync|createReadStream|createWriteStream|readdir|readdirSync|rm|rmSync|unlink|unlinkSync|open|openSync)\s*\([^)]*\breq\s*\.\s*(?:params|query|body|headers|url|originalUrl)\b@: fs.* sink references req.*. \texttt{bundle}.
\end{itemize}

\paragraph{Prototype Pollution}
\begin{itemize}[label=--]
  \item \verb@\bObject\s*\.\s*assign\s*\([^)]*\b(?:req\s*\.\s*(?:body|query|params)|JSON\s*\.\s*parse\s*\(|qs\s*\.\s*parse\s*\()\b[^)]*\)@: Object.assign fed by req.* or JSON.parse.
  \item \verb@=\s*\{[^}]*\.\.\.(?:req\s*\.\s*(?:body|query|params)|JSON\s*\.\s*parse\s*\(|qs\s*\.\s*parse\s*\()[^}]*\}@: Spread merge with attacker-controlled object.
  \item \verb@\b(?:set|assign|merge|extend|defaultsDeep|deepMerge|deepExtend)\b\s*\([^)]*\b(?:req\s*\.\s*(?:body|query|params)|JSON\s*\.\s*parse\s*\(|qs\s*\.\s*parse\s*\()\b@: Generic deep merge helpers with tainted input.
  \item \verb@\b_\s*\.\s*(?:merge|mergeWith|defaultsDeep|set|setWith|update|updateWith)\s*\(@: lodash-style risky helpers.
\end{itemize}

\section{Sampling Algorithms for Crawled \npm Packages}\label{appendix:sampling}

We characterize each package using four features: three structural metrics
$\mathtt{js\_ts\_loc}$ (lines of JavaScript or TypeScript code),
$\mathtt{dependency\_count}$ (number of declared dependencies), and
$\mathtt{js\_ts\_files}$ (number of JavaScript or TypeScript source files),
together with a binary indicator of whether the package contains minified code (by simply checking whether \texttt{.min.} or \texttt{bundle} appears in the filenames of JavaScript or TypeScript files).
Our goal is to sample a fixed number of packages per vulnerability (65 in our experiments) such that the distribution over the three structural metrics is approximately uniform, while the proportion of minified packages follows the distribution observed in the crawled dataset, subject to a minimum of one minified package whenever minified artifacts exist.

We implement this goal using a stratified sampling procedure:
\begin{itemize}
  \item \textbf{Metric Bucketing}: For each vulnerability class $v$ and each structural metric
  $m_j \in \{\allowbreak\mathtt{js\_ts\_loc},\allowbreak \mathtt{dependency\_count},\allowbreak \mathtt{js\_ts\_files}\}$,
  we discretize metric values into $b=5$ buckets using empirical quantiles.
  For $k \in \{1,\dots,b-1\}$, we define bucket cutoffs as
  $\theta_{j,k} = \text{quantile}_{k/b}(\{m_j(x)\mid x\in v\})$,
  and assign each package $x$ to a bucket $B_j(x)$ according to the interval in which
  $m_j(x)$ falls.

  \item \textbf{Group Definition}: Each package is assigned to a joint group
  \[
    S(x) = (B_1(x), B_2(x), B_3(x), M(x)),
  \]
  where $M(x)\in\{\text{minified}, \text{plain}\}$ denotes the minification status.

  \item \textbf{Structural Allocation}: Ignoring the minification indicator, we target an approximately uniform allocation across the structural bucket tuples $(B_1, B_2, B_3)$ by iterating these tuples in randomized round robin order and selecting packages to spread samples evenly across different structural configurations.

  \item \textbf{Minification Allocation}: Let $C_{\text{min}}$ and $C_{\text{plain}}$ denote the counts of minified and plain packages in the crawled dataset for vulnerability $v$.
  We set sampling budgets $N_{\text{min}}$ and $N_{\text{plain}}$ proportional to $C_{\text{min}}$ and $C_{\text{plain}}$, while enforcing $N_{\text{min}} \ge 1$ whenever $C_{\text{min}} > 0$.

  \item \textbf{Selection}: Within each structural bucket tuple $(B_1, B_2, B_3)$, we draw packages while respecting the remaining minified and plain budgets. If one category is not available within a bucket, we draw from the other category. If a quota cannot be met due to global exhaustion, we fill the remaining slots from the available category.

  \item \textbf{Post Check}:
  For each metric $m_j$ and each bucket $t$ that appears among the sampled packages, if the sample contains at least one plain package in $(m_j,t)$ but zero minified packages in $(m_j,t)$, and the crawled dataset contains at least one minified package in $(m_j,t)$, then we replace one sampled plain package from $(m_j,t)$ with an unused minified package from $(m_j,t)$ if such a minified package exists.
\end{itemize}

\section{Prompts Templates}\label{appendix:prompts}
Here we include the detailed prompt templates used for each stage of the pipeline and for each vulnerability class. The placeholder variables in the templates are dynamically filled during the execution of the pipeline.

\subsection{Stage 1: Finder Prompt}
\subsubsection{System Prompt}
\begin{verbatim}
You are an expert security researcher specializing in Node.js vulnerabilities.

Your task is to find instances of <VULN_TYPE> vulnerabilities in the project at: <PROJECT_PATH>

<VULN_DESCRIPTION>

WORKFLOW:
1. Start by getting the file tree or listing files to understand the structure
2. Search for patterns related to <VULN_TYPE>
3. Read suspicious files to analyze the code
4. Identify exact locations (file + line number) of vulnerabilities
5. Determine which public APIs can reach these vulnerabilities
6. Call submit_findings(findings=[...]) with structured arguments (NO JSON STRINGS). You can call this multiple times as you discover items.
7. When you are completely done adding findings, call finish(summary="...optional...") to end the run.

FINDINGS FORMAT (submit_findings arguments):
- findings: [
        {
            "vuln_type": "<VULN_TYPE>",
            "file": "relative/path/to/file.js",
            "line": 42,
            "description": "Brief description",
            "evidence": "Code snippet showing the issue",
            "reachable_apis": ["api1", "api2"],
            "confidence": 0.85
        }
    ]

Be thorough and precise. Focus on actionable evidence but err on the side of inclusion: if a spot looks plausibly exploitable yet you lack full confirmation, include it with a lower confidence score and clearly state any assumptions. When you have submitted all findings, call finish to end the run.
\end{verbatim}

\subsubsection{User Prompt}
\begin{verbatim}
Find all <VULN_TYPE> vulnerabilities in the project. Use tools to analyze the code, then submit your findings.
\end{verbatim}

\subsection{Stage 2: Judge Prompt}
\subsubsection{System Prompt}
\begin{verbatim}
You are an expert security code reviewer specializing in Node.js vulnerabilities.

Your task is to validate whether a reported <VULN_TYPE> vulnerability is actually exploitable.

CURRENT DIRECTORY: <PROJECT_PATH>
All file paths are relative to this directory. Call get_file_tree() first to see the structure if needed.

ANALYSIS CHECKLIST:
1. Read the code at the reported location
2. Trace data flow to see if user input can reach the vulnerable sink
3. Check for any input validation or sanitization
4. Determine if the vulnerability is actually exploitable
5. Submit your verdict with detailed reasoning

IMPORTANT - LIBRARY/PACKAGE ATTACK SURFACE:
- When analyzing npm packages/libraries, EXPORTED functions (exports.*, module.exports) are the attack surface
- If a vulnerable function is exported (even with no callers in the codebase), it IS reachable by external code
- Focus on: Can user-controlled input reach the sink IF the exported function is called?
- Do NOT search for external callers - the export itself makes it callable

VERDICT FORMAT (submit_verdict arguments):
- is_valid: true/false
- reason: "Detailed explanation with evidence"
- confidence: 0.0-1.0

Call submit_verdict(is_valid=true/false, reason="...", confidence=0.0-1.0). Do NOT wrap the arguments in JSON strings.

Be conservative - only mark as valid if truly exploitable.
\end{verbatim}

\subsubsection{User Prompt}
\begin{verbatim}
Validate this vulnerability finding:

Type: <VULN_TYPE>
File: <FILE>
Line: <LINE>
Description: <DESCRIPTION>
Evidence: <EVIDENCE>
Reachable APIs: <REACHABLE_APIS>
Confidence: <CONFIDENCE>

Use tools to analyze the code and submit your verdict.
\end{verbatim}

\subsection{Stage 3: Constraints Inferencer Prompt}
\subsubsection{System Prompt}
\begin{verbatim}
You are an expert security researcher analyzing how to exploit vulnerabilities.

Your task is to analyze a <VULN_TYPE> vulnerability and determine the constraints for exploitation.

ANALYSIS OBJECTIVES:
1. Identify the entry point (URL, function, API)
2. Determine parameter names and how to provide them
3. Understand required format/structure of the payload
4. Identify any validation to bypass
5. Define success criteria for the exploit

CONSTRAINTS FORMAT (submit_constraints arguments):
- constraints: "Detailed natural language description of exploitation requirements" (REQUIRED)
- entry_point: "How to reach the vulnerability"
- parameters: ["param1", "param2"]
- payload_format: "Required format for the payload"

Be specific and actionable - the exploit agent will use this.
\end{verbatim}

\subsubsection{User Prompt}
\begin{verbatim}
Generate exploitation constraints for:

Type: <VULN_TYPE>
File: <FILE>
Line: <LINE>
Description: <DESCRIPTION>
Evidence: <EVIDENCE>

Analyze the code and submit detailed constraints.
\end{verbatim}

\subsection{Stage 4: Exploiter Prompt}
\subsubsection{System Prompt}
\begin{verbatim}
You are an expert security researcher crafting exploits for <VULN_TYPE>.

Your task is to create a working exploit based on the provided constraints.

EXPLOIT GOAL:
<EXPLOIT_INSTRUCTIONS_FROM_VULN_DEFINITION>

Dependency setup: Install any missing dependencies in the package directory before running your exploit (use run_shell_command if you need npm/yarn).

EXPLOIT WORKFLOW:
1. Read the vulnerable code to understand it
2. Craft exploit code that imports the module: const mod = require('./index');
3. Execute the exploit using execute_javascript
   - Only if you truly need a background service (e.g., start a server), use start_persistent_process, inspect with check_persistent_process, and always call kill_persistent_process when finished.
4. Verify success (check file creation, output, etc.)
5. If import fails, check the exact filename and try again
6. Submit your result with submit_exploit_result(success=..., exploit_code="...", execution_output="...", explanation="...")

Be creative and persistent. Learn from failures.
\end{verbatim}

\subsubsection{User Prompt}
\begin{verbatim}
Create and execute an exploit for:

Type: <VULN_TYPE>
File: <FILE>
Line: <LINE>

Constraints:
<CONSTRAINTS_TEXT>

Create a working exploit and submit the result.
\end{verbatim}

\subsection{Stage 4 Dynamic Goal Templates (Per Vulnerability Type)}
\subsubsection{os\_command\_injection}
\begin{verbatim}
Craft a payload that injects OS commands (e.g., using ; || && or backticks).
The exploit should create the file /tmp/os_cmd_success to prove arbitrary command execution.
Example: `; touch /tmp/os_cmd_success` or `$(touch /tmp/os_cmd_success)`
\end{verbatim}

\subsubsection{code\_injection}
\begin{verbatim}
Craft a payload that injects JavaScript code into eval(), Function(), vm.runInNewContext(),
or similar APIs. The exploit harness defines global.CTF = () => console.log('GLOBAL.CTF HIT').
Your malicious payload must call global.CTF() so that the marker appears in STDOUT, proving
arbitrary code execution without relying on filesystem writes.
\end{verbatim}

\subsubsection{path\_traversal}
\begin{verbatim}
Craft a payload using ../ sequences to escape the intended directory and read /tmp/path_traversal.
The harness writes the sentinel string 'PATH_TRAVERSAL' into that file ahead of time.
Use the vulnerability to read it and console.log the file contents so STDOUT contains the marker.
\end{verbatim}

\subsubsection{prototype\_pollution}
\begin{verbatim}
Goal: trigger prototype pollution so that a newly created object inherits attacker-controlled data.

Use the package's public API as intended (e.g., merge, copy, set, or path-based helpers).
The attacker controls input values such as objects, keys, or paths.

You may use payloads such as:
- objects containing prototype-related keys
- dotted paths (e.g., '__proto__.polluted', 'constructor.prototype.polluted')
- arrays of path segments if supported by the API

Do NOT use unrealistic techniques such as Object.defineProperty, Object.setPrototypeOf, or
direct writes to Object.prototype.

IMPORTANT:
- Do NOT try to manually check for success.
- The JavaScript run tool automatically probes, and if pollution occurred it will emit
'PROTO_POLLUTION SUCCESS' on its own.
- Do NOT print or hardcode the success token.
\end{verbatim}

\subsection{Vulnerability Definitions}
\subsubsection{os\_command\_injection}
\begin{verbatim}
CWEs: 077, 078

OS Command Injection (CWE-78): The product constructs and executes OS commands
using unsanitized input, allowing attackers to execute arbitrary system commands.

Common sink APIs: exec, execSync, execFile, execFileSync, spawn, spawnSync,
child_process module functions
\end{verbatim}

\subsubsection{code\_injection}
\begin{verbatim}
CWEs: 094

Code Injection (CWE-94): The product dynamically generates or evaluates code
using untrusted input, allowing attackers to inject and execute arbitrary JavaScript code.
\end{verbatim}

\subsubsection{path\_traversal}
\begin{verbatim}
CWEs: 022, 035

Path Traversal (CWE-22): The product uses external input to construct a file or directory
path but fails to properly sanitize special characters (e.g., ../), allowing access to
unintended locations outside the intended directory.
\end{verbatim}

\subsubsection{prototype\_pollution}
\begin{verbatim}
CWEs: 1321

Prototype Pollution (CWE-1321): attacker-controlled keys or paths are used in object writes or
merge/copy operations, causing a shared prototype (often Object.prototype) to be modified and
affecting subsequently created objects.

Indicators include dynamic property access (obj[key]), merge/copy utilities, recursive assignment
patterns, and path-based setters. The vulnerable code may not explicitly reference '__proto__' or
'constructor.prototype'.
\end{verbatim}

\section{Full File-Level Matching Results and Metric Definitions}
\label{app:unmatched}

\begin{table*}[t]
\centering
\setlength{\tabcolsep}{4pt}
\caption{Detailed file-level unmatched results by dataset and vulnerability type for Finder and Judge stages.
``Findings Unmatched'' reports unmatched-rate on tool findings, and ``GT Unmatched'' reports unmatched-rate on ground-truth files.}
\label{tab:file-level-unmatched-summary}
\begin{tabular}{llcccc}
\hline
Dataset & Vulnerability & Finder Findings Unmatched & Judge Findings Unmatched & Finder GT Unmatched & Judge GT Unmatched \\
\hline
SecBench.js & CWE-22 & 27/221 (12.22\%) & 25/218 (11.47\%) & 2/156 (1.28\%) & 2/156 (1.28\%) \\
SecBench.js & CWE-471 & 113/343 (32.94\%) & 92/299 (30.77\%) & 10/118 (8.47\%) & 12/118 (10.17\%) \\
SecBench.js & CWE-78 & 21/201 (10.45\%) & 18/194 (9.28\%) & 3/78 (3.85\%) & 3/78 (3.85\%) \\
SecBench.js & CWE-94 & 28/59 (47.46\%) & 25/55 (45.45\%) & 1/21 (4.76\%) & 1/21 (4.76\%) \\
SecBench.js & All & 189/824 (22.94\%) & 160/766 (20.89\%) & 16/373 (4.29\%) & 18/373 (4.83\%) \\
\hline
VulcaN & CWE-22 & 4/8 (50.00\%) & 4/8 (50.00\%) & 0/3 (0.00\%) & 0/3 (0.00\%) \\
VulcaN & CWE-471 & 34/146 (23.29\%) & 26/130 (20.00\%) & 6/63 (9.52\%) & 6/63 (9.52\%) \\
VulcaN & CWE-78 & 40/163 (24.54\%) & 35/128 (27.34\%) & 5/63 (7.94\%) & 8/63 (12.70\%) \\
VulcaN & CWE-94 & 30/58 (51.72\%) & 28/53 (52.83\%) & 5/22 (22.73\%) & 8/22 (36.36\%) \\
VulcaN & All & 108/375 (28.80\%) & 93/319 (29.15\%) & 16/151 (10.60\%) & 22/151 (14.57\%) \\
\hline
All & CWE-22 & 31/229 (13.54\%) & 29/226 (12.83\%) & 2/159 (1.26\%) & 2/159 (1.26\%) \\
All & CWE-471 & 147/489 (30.06\%) & 118/429 (27.51\%) & 16/181 (8.84\%) & 18/181 (9.94\%) \\
All & CWE-78 & 61/364 (16.76\%) & 53/322 (16.46\%) & 8/141 (5.67\%) & 11/141 (7.80\%) \\
All & CWE-94 & 58/117 (49.57\%) & 53/108 (49.07\%) & 6/43 (13.95\%) & 9/43 (20.93\%) \\
All & All & 297/1199 (24.77\%) & 253/1085 (23.32\%) & 32/524 (6.11\%) & 40/524 (7.63\%) \\
\hline
\end{tabular}
\end{table*}

The metrics we used for file-level matching are formally defined as follows: let $d$ denote a dataset (SecBench.js, VulcaN, or their union) and $s \in \{\text{Finder},\text{Judge}\}$ denote a stage. For each $(d,s)$:
\begin{itemize}[leftmargin=*,itemsep=2pt,topsep=2pt]
	\item Let $F_{d,\text{Finder}}$ be the set of findings produced by the Finder stage, and let $F_{d,\text{Judge}}$ be the subset of findings deemed valid by the Judge stage.
	\item Let $G_d$ be the set of ground-truth vulnerable files.
	\item A finding $f \in F_{d,s}$ is \emph{matched} if there exists $g \in G_d$ under our file-level matching rule (same package-version pair, vulnerability type, and file). Otherwise, $f$ is \emph{unmatched}.
\end{itemize}

We report the following four metrics, which correspond directly to \autoref{tab:file-level-unmatched-small} and \autoref{tab:file-level-unmatched-summary}:
\begin{align*}
\text{Finder Findings }&\text{Unmatched}(d) \\
&= \frac{|\{f \in F_{d,\text{Finder}}: f\text{ unmatched}\}|}{|F_{d,\text{Finder}}|} \\
\text{Judge Findings }&\text{Unmatched}(d) \\
&= \frac{|\{f \in F_{d,\text{Judge}}: f\text{ unmatched}\}|}{|F_{d,\text{Judge}}|} \\
\text{Finder GT }&\text{Unmatched}(d) \\
&= \frac{|\{g \in G_d: \nexists f \in F_{d,\text{Finder}}\text{ matched to }g\}|}{|G_d|} \\
\text{Judge GT }&\text{Unmatched}(d) \\
&= \frac{|\{g \in G_d: \nexists f \in F_{d,\text{Judge}}\text{ matched to }g\}|}{|G_d|}
\end{align*}

The first two metrics measure the fraction of tool reports that are unmatched; the latter two measure the fraction of benchmark vulnerable files not covered by any report at each stage.

As shown in Table~\ref{tab:file-level-unmatched-summary}, the detailed file-level matching results are broken down by benchmark (SecBench.js and VulcaN) and vulnerability type, including both unmatched-finding rates and unmatched-ground-truth rates.

\end{document}